# Defect Dipole Induced Improved Electrocaloric Effect in Modified NBT-6BT Lead-Free Ceramics

Koyal Suman Samantaray[1], Ruhul Amin[1], E.G Rini[1], Indranil Bhaumik[2,3], A. Mekki[4,5], K. Harrabi[4], Somaditya Sen[1]

[1]Department of Physics, Indian Institute of Technology Indore, Indore, 453552, India

[2]Crystal Growth and Instrumentation Section, Laser and Functional Materials Division, Raja Ramanna Centre for Advanced Technology, Indore- 452013, India

[3]Homi Bhabha National Institute, Training School Complex, Anushakti Nagar, Mumbai-400094, India

[4]Department of Physics, King Fahd University of Petroleum & Minerals Dhahran, 31261, Saudi Arabia

[5]Center for Advanced Material, King Fahd University of Petroleum & Minerals, Dhahran, 31261, Saudi Arabia

ABSTRACT:

The Rietveld refinement of the polycrystalline powders of 1% Fe and Mn-doped $(Na_{0.5}Bi_{0.5})_{0.94}Ba_{0.06}Ti_{0.98}V_{0.02}O_3$ at the Ti-site confirmed a single rhombohedral (*R3c*) phase. The bandgap, ($E_g$) was affected by the anti-phase octahedral tilt angle and the spin-orbit splitting energy of $Ti^{4+}2p_{3/2}$ and $Ti^{4+}2p_{1/2}$ states. The decrease in Bi loss and increase in the binding energy of Ba due to Fe/Mn doping has been correlated to the strengthening of Bi-O and Ba-O bonds which was revealed from the XPS studies thereby further related to the average A-O bond length from structural studies. Hence, a reduction of oxygen vacancy ($V_O$) for the doped samples has been justified. A significant improvement of the dielectric constant, relaxation time ($\tau_0$), and the decrease in conductivity due to doping was revealed from the frequency-dependent (10Hz-1MHz) dielectric measurement study. The conduction and relaxation process is dominated by the short-range movement of defects. The activation energy ($E_a$ ~1eV) revealed that there is a presence of double-ionized $V_O$s. The ECE study showed a significant enhancement of the changes in entropy, $\Delta S$, and the adiabatic temperature difference, $\Delta T$, due to doping, with $\Delta T$ being highest in the Fe-doped sample. Such improvement of dielectric and ECE properties was confirmed due to the reduction of the mobility of oxygen vacancy because of the formation $(Mn''_{Ti} - V_o^{\bullet\bullet})/(Mn'_{Ti} - V_o^{\bullet\bullet})^{\bullet}$ and $Fe''_{Ti} - V_o^{\bullet\bullet}/(Fe'_{Ti} - V_o^{\bullet\bullet})^{\bullet}$ defect dipoles.

INTRODUCTION:

Perovskite oxides with dielectric, ferroelectric, and piezoelectric properties display structural phase transitions providing a scope of interesting physics. PZT is an important member of the family of perovskite oxides due to its wide applications[1,2]. But Pb being a toxic element has been restricted and banned globally which has affected the use of PZT in industries[3]. Hence, Pb-free materials are evolving today in different energy storage applications[4,5]. $Na_{0.5}Bi_{0.5}TiO_3$ (NBT) is the foremost leader in the field with excellent electrical properties [6–8]. However, a high leakage current and high coercivity in NBT is a problem that degrades its energy storage capabilities [9,10]. The leakage properties arise from the evaporation of volatile elements (Na and/or Bi) which results in oxygen vacancies, $V_O$ and degrades the energy storage capabilities [11]. Hence, various chemical modifications have been explored [12,13]. A handful of new materials like NBT-BT, NBT-KNN. NBT-ST etc. has been invented with great potential [13–15]. NBT6BT has also been modified for achieving different high-end applicability [16,17].

One of the growing demands is in the high electrocaloric (ECE) properties of materials [18,19]. The adiabatic change in temperature, $\Delta T$, is the crucial factor that determines a material's applicability in refrigerators and cooling devices [20,21]. PZT thin films have a giant EC effect (ECE) with $\Delta T \sim 12K$, making PZT the most efficient material for refrigeration and solid-state cooling applications [22]. Among Pb-free materials, NBT-based materials are gaining importance in the ECE field. Zhang et. al reported a negative. $\Delta T = -1$ K at for an applied electric field of 60 kV/cm at a temperature of $T = 50–70\,°C$ in NBT-6BT [23]. Due to the presence of defect dipoles the local ordered polarization regions around it become relatively disordered due to the application of the external electric field, hence improvement in the value of $\Delta S$ [24,25]. A series of reports followed on A/B site modified NBT-6BT systems, stating enhanced ECE [23,26]. By defect engineering, ECE can be tuned to a great extent as the defect dipoles are quite sensitive to the applied electric field [24,27,28].

V-doped NBT-6BT series was reported showing V-doping showing enhanced dielectric and ferroelectric properties [11]. The V doped sample was further modified by substituting Fe in place of Ti showing 1% Fe-doping showing a decrease in coercive field and enhancement in remnant polarization [29]. Fe generally prefers to be in the $Fe^{3+}$ state. However, the valence state was not confirmed and ECE was not studied in that report. Along with S0 and Fe-S0, a comparative study of a similar 1% Mn-doped S0 system is reported in this study on a series with the general formula of $(Na_{0.5}Bi_{0.5})_{0.94}Ba0_{.06}Ti_{0.97}V_{0.02}M_{0.01}O_3$ (where S0 refers to M= Ti, Fe-S0 refers to M=Fe and Mn-S0 refers to M=Mn). A detailed structural, optical, valence state, dielectric, and electrocaloric study of the polycrystalline powders are being reported for the first time to understand the importance of valence state and ionic sizes of the dopants on the structure and hence the physical properties. The rhombohedral, hexagonal and pseudo cubic representations of the *R3c* space group are explored in these three samples. The presence of $Mn_{Ti}'' - V_o^{\bullet\bullet} / Fe_{Ti}'' - V_o^{\bullet\bullet}$ defect dipoles in the samples are investigated. A study of the complex impedance spectroscopy, modulus spectroscopy, and AC conductivity analysis was performed to understand the relaxation and conduction mechanism in these samples. Finally, the influence of defect dipoles in enhancement of the ECE of the materials is discussed.

METHODOLOGY:

The polycrystalline powders of S0, Fe-S0, and Mn-S0 were synthesized using a process involving a sol-gel synthesis followed by stepwise ambient sintering at high temperatures at 700ºC (8h) and thereafter at 1100ºC (2h). The phase was verified from x-ray diffraction (XRD) using an x-ray diffractometer (Bruker D2-Phaser). The XRD pattern of all the samples was refined using Rietveld refinement. The refinement was done using the FullProf suite software considering pseudo-Voigt peak shape for all the samples. The bandgap data was collected using an Ocean Optics UV-Vis spectrometer.

The chemical composition and the valence states of the elements present in the prepared samples were studied using the X-Ray photoelectron spectroscopy experiment which was performed using the Thermo-Scientific Escalab 250 Xi XPS Spectrometer (Al-Kα x-rays) having an energy resolution of ~ 0.5 eV. The spectra were deconvoluted using XPSPEAK41 software. The XPS spectra were initially calibrated using the C1s peak. The background was properly extracted using the Tougaard function. A combined Gaussian-Lorentzian peak shape was used for all the peaks to get quality fitting.

The electronic characterization for all the samples was done on the sintered pellets of the final phase polycrystals at 1100°C for 2hr. For dielectric measurement, silver electrodes were prepared on both sides of the pellets using silver paste. The pellets were thereafter annealed for proper adhesion of the Ag to the pellets at 540ºC for 15 minutes. The dielectric measurement was done using a broadband dielectric spectrometer using Newton's $4^{th}$ Ltd. phase-sensitive multimeter having signal strength $1V_{rms}$ in the temperature range 50ºC to 450ºC and frequency range 1Hz-1MHz. A ferroelectric (P-E) loop tracer instrument (M/s Radiant Instruments, USA) was used to investigate the ferroelectric properties of the same pellets. The temperature-dependent PE loop for 35-130ºC temperature range was done to study the electrocaloric effect.

RESULTS AND DISCUSSIONS:

X-Ray Diffraction Study:

The goodness of fit obtained for all the samples was within the acceptable limits [Fig.1]. For the *R3c* space group, two kinds of axes preference can be used. One is hexagonal axes ($a_H$, $c_H$) and the other one is rhombohedral axes ($a_R$). Fig.2 (a), (b) depicts the unit cell presentation in hexagonal and rhombohedral settings, respectively. *R3c* hexagonal settings are usually used for the NBT based samples as this provides information about the octahedral tilt angle, and octahedral distortion easily from the atomic positions [17,30]. The lattice parameters of *R3c* can be converted from hexagonal settings ($a_H$, $c_H$, $\alpha=\beta=90º$, and $\gamma= 120º$) to rhombohedral settings ($a_R$, $\alpha_R$) and vice-versa using the following expressions. The rhombohedral lattice parameter can be calculated from hexagonal settings using the expressions: $a_R = a_H \frac{\sqrt{3+(\frac{c_H}{a_H})^2}}{3}$, $\alpha_R = cos^{-1}(\frac{2c_H^2 - 3a_H^2}{2c_H^2 + 6a_H^2})$. Similarly, the hexagonal lattice parameters can be calculated from the rhombohedral settings using $a_H = 2a_R \sin(\frac{a_R}{2})$, and $c_H = a_R 3\sqrt{(1 + 2\cos a_R)}$ . Any perovskite unit cell belonging to various structural space groups can be converted to an equivalent pseudo-cubic unit cell for a suitable comparison. The rhombohedral axes fall on the face diagonal of the pseudo-cubic unit cell. So, the rhombohedral unit cell parameters ($a_R$) can be converted to the pseudo-cubic cell parameters ($a_p$) using the relation: $a_P = \frac{a_R}{\sqrt{2}}$ [31]. The values of lattice parameters of all the three samples in rhombohedral, hexagonal, and pseudo-cubic settings using the above-discussed relations are tabulated in Table I.

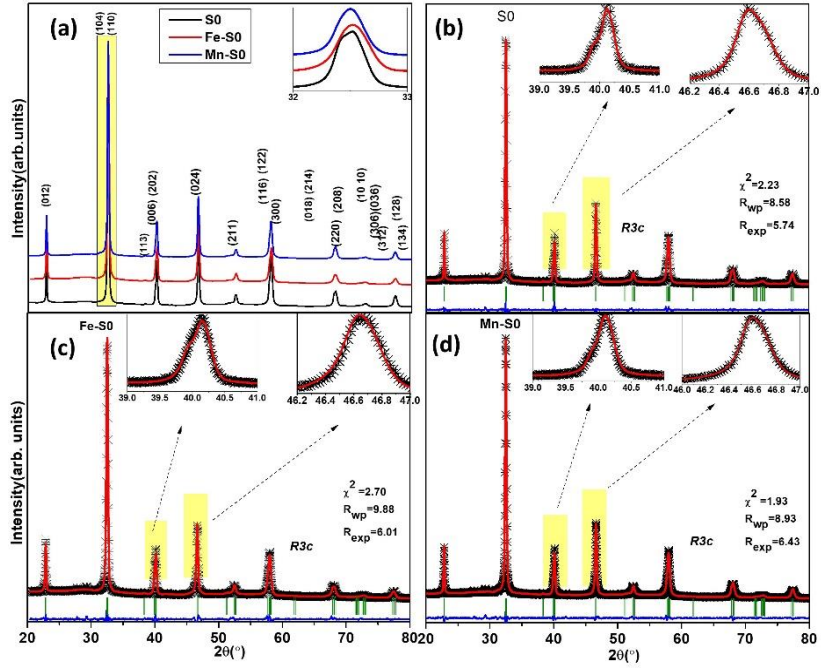

**Figure 1** **(a)** Comparison of XRD pattern of all the three compositions and Rietveld refinement plots for **(b)** S0 **(c)** Fe-S0 **(d)** Mn-S0 compositions

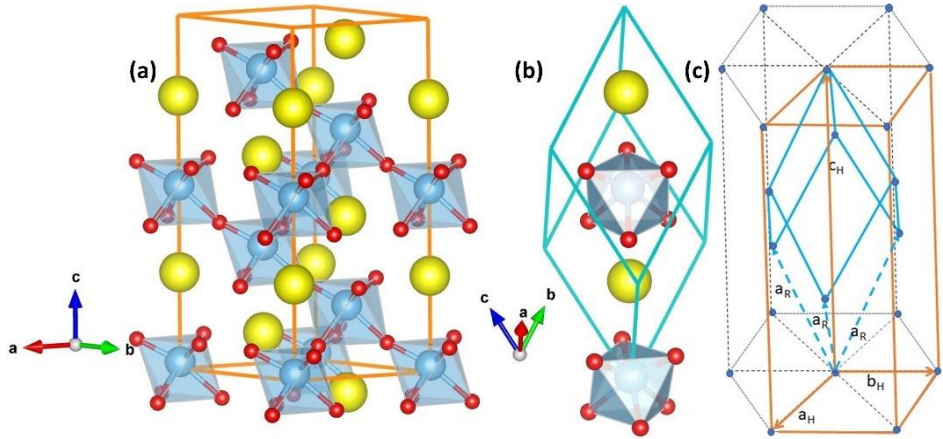

**Figure 2** **(a)** Representation of NBT lattice in hexagonal settings **(b)** Representation of NBT lattice in rhombohedral settings **(c)** Diagram depicting the relation between hexagonal and rhombohedral lattice representations.

**Table I** Lattice parameters of three compositions for the lattice in hexagonal, rhombohedral, and pseudocubic settings

| Compositions | Lattice Parameters | | | | |
|---|---|---|---|---|---|
| | Hexagonal | | Rhombohedral | | Pseudo-cubic |
| | $a_H=b_H$ (Å) | $c_H$ (Å) | $a_R=b_R=c_R$ (Å) | $\alpha_R$ (°) | $a_p$ (Å) |
| Fe-S0 | 5.4964(4) | 13.5463(2) | 5.5190(1) | 59.7294(3) | 3.9025(2) |
| S0 | 5.5005(6) | 13.5534(2) | 5.5223(1) | 59.7391(2) | 3.9048(5) |

| Mn-S0 | 5.5041(2) | 13.5526(4) | 5.5232(8) | 59.7105(7) | 3.9055(5) |

A contraction is observed in the lattice parameters in the case of the Fe-S0 sample while expansion is observed in the Mn-S0 sample as compared to the S0 sample [Table I]. This may be related to the ionic size of the constituent B-site ions as there are various possibilities of the ionic valence states of Ti, V, Fe, and Mn in these materials. A Shannon radii study of such possibilities reveal the following results for hexa-coordinated ions: $Ti^{4+}$ ~0.745 Å, $Fe^{2+}$(low spin) ~0.75 Å, $Fe^{2+}$(high spin) ~0.92 Å, $Fe^{3+}$(low spin) ~0.69 Å, $Fe^{3+}$(high spin) ~0.785 Å, $Mn^{2+}$(low spin) ~0.81 Å, $Mn^{2+}$(high spin) ~0.97 Å, $Mn^{3+}$(low spin) ~0.72 Å, $Mn^{3+}$(high spin) ~0.785 Å, $Mn^{4+}$ ~0.67 Å, $V^{3+}$ ~0.78 Å, $V^{4+}$ ~0.72 Å and $V^{5+}$ ~0.68 Å. On the other hand, the A-site ions of NBT can have only two possibilities: $Na^+$(XII) ~1.53 Å, and $Ba^{2+}$(XII) ~1.75 Å. An option of Bi ionic radii for XII coordination is not available. The values available for Bi ions are for $Bi^{3+}$(VIII) ~1.31 Å, $Bi^{5+}$(VI) ~0.9 Å. However, a rough estimation by extrapolation method may suggest Bi in XII coordination to assume a size of approximately 1.5 Å and 1.3 Å for $Bi^{3+}$ and $Bi^{5+}$ respectively. When $Bi^{3+}$ is of the order of $Na^+$ ions, $Bi^{5+}$ is much smaller. Hence, an ionic valence state study using XPS becomes extremely important in this report.

The above comparison of the lattice in all three representations reveals an interesting fact. While the pseudocubic and the rhombohedral cell lattice parameters increase in the order of Fe-S0 → S0 → Mn-S0, the hexagonal cell lattice parameters show two different trends [Table I]. While the a, b axes increase in the same trend, the c-axis is the largest for S0 and least for Fe-S0. These effects may be due to the different distortions in the $BO_6$ octahedral structure which appears to have two sets of three equal B-O bonds. As a result, the tilting angle of these octahedra also varies. The tilting angle of the Fe-S0 sample is ~3.71º, while for S0 it is ~4.88º and for Mn-S0 it is 5.04º [30]. Hence, the tilting angle increases in the same order as Fe-S0 → S0 → Mn-S0, like the pseudocubic lattice parameters.

**Table II** Average bond lengths and bond angles calculated in Mercury software using the CIF files after refinement

| Average Bond Length and Bond Angle | Fe-S0 (A) | S0 (A) | Mn-S0 (A) |
|---|---|---|---|
| Large B-O (Å) | 2.144 | 1.963 | 2.094 |
| Short B-O (Å) | 1.835 | 1.963 | 1.843 |
| Short <O-B-O> (º) | 74.07 | 81.56 | 80.69 |

| Large <O-B-O> (º) | 105.06 | 97.69 | 99.66 |
|---|---|---|---|
| A$_1$-O (Å) | 3.047 | 3.028 | 3.035 |
| A$_2$-O (Å) | 2.662 | 2.899 | 2.630 |
| A$_3$-O (Å) | 2.479 | 2.502 | 2.503 |
| A$_4$-O (Å) | 2.867 | 2.628 | 2.909 |
| | | | |
| <O-A-O> (º) | 62.81 | 61.66 | 65.53 |
| <O-A-O> (º) | 57.38 | 58.44 | 55.28 |
| Opp.<O-B-O> (º) | 157.06 | 168.55 | 166.42 |

Optical Bandgap studies:

The optical band gap energy, E$_g$, is 3.35eV for Fe-S0 which is slightly higher than S0 with ~3.33eV. With the Mn-doping the value of E$_g$ reduced further to ~2.90eV for Mn-S0 [Fig.S1]. This variation of E$_g$ can be better explained from the variation in octahedral tilt angle (from XRD studies) and spin-orbit splitting energy of the Ti$^{4+}$2p$_{3/2}$ and Ti$^{4+}$2p$_{1/2}$ (from XPS studies). The overlap of Ti 2p and O 1s orbital is modified due to the spin-orbit splitting of the Ti$^{4+}$ 2p orbitals (XPS studies). The Urbach energy on the other hand seems to continuously increase from ~0.275eV (Fe-S0) to ~0.282 eV (S0) and 0.330 eV in Mn-S0 [Fig.S1]. Hence, the trend observed in these samples must have some correlation with the valence state of the constituent ions which have naturally been modified to accommodate the given stoichiometry maintaining a single-phase or structure.

XPS Analysis:

i.     O 1s core-level spectra:

The oxygen content in oxide is one of the driving forces behind the different structure correlated properties of a material. The O-1s spectra were deconvoluted to three peaks which lie in the range of 529–530 eV, 530–532 eV, and 533–534 eV corresponding to the binding energy of lattice oxygen (O$_L$), oxygen vacancy (O$_V$), and adsorbed oxygen (O$_A$), respectively [32–35]. The O$_A$ is located only at the surface and is a dynamic entity depending on the external parameters. Hence, considering only the O$_L$ and O$_V$ contributing to the lattice, a numerical estimation yields an O$_V$ fraction of 0.39 for Fe-S0, 0.62 for S0, and 0.25 for Mn-S0 [Fig.5(d)-(f)]. Hence, the incorporation of Fe and Mn helps to reduce O$_V$, with Mn being the better of the two. This is an indication of probable cation-loss in the structure for the S0 sample larger than the doped samples. The variation of O$_V$ is due to the systematic change in the A-site cationic contribution in the respective samples. Due to the high-temperature sintering of NBT based samples, the lattice loses the volatile Na/Bi ions which can provide a possible explanation for the anomaly [36]. The Fe/Mn doping, therefore, helps in A-site ion-loss and thereby reduces the creation of $O_V$. On the other hand, the fraction of $O_A$ concerning the lattice and vacant oxygen sites was found to increase from 0.10, Fe-S0, to 0.13 for S0, and 0.42 Mn-S0, respectively. The adsorbed oxygen is mostly due to the uncompensated surface charges which may be due to the changes in the net cationic charges in the materials. Hence, a valence state study of the individual cations becomes extremely important. All the observed binding energy and FWHM are tabulated in Table III.

ii.     Na 1$s$ core-level spectra:

The core Na 1$s$ peak confirmed $Na^+$ state for all the samples [Fig. 3(a)-(c)]. Extra peaks observed in this region at ~1070.59 eV and ~1074.26 eV which indicate the presence of two Ti-LMM Auger transitions and will be discussed in the following sections [37–39]. The presence of two Ti-LMM Auger peaks along with the Na 1$s$ peak was observed for all the samples.

iii.    Bi 4$f$ core-level spectra:

A mixed oxidation state of $Bi^{3+}$ and $Bi^{5+}$ is observed for all the samples [Fig. 3(d)-(f)]. The $Bi^{3+}$ 4$f_{7/2}$ and $Bi^{3+}$ 4$f_{5/2}$ peak is observed at (160.26 eV, 165.51 eV) for Fe-S0, (159.40 eV, 164.60 eV) for S0 and (159.86 eV, 165.17 eV) for Mn-S0 [40–42]. The spin-orbit splitting energy for $Bi^{3+}$4$f_{7/2}$ and $Bi^{3+}$4$f_{5/2}$ was observed to be continuously increased from 5.20 eV for S0 to 5.25 eV for Fe-S0 and 5.31 eV for Mn-S0 samples. The increase in B.E for Fe-S0 and Mn-S0 samples is due to the increased contribution of $Bi^{3+}$ to the desired A-site position in the lattice which consequently increased the lattice oxygen in the materials. That means the Bi-O bonding is becoming stronger with Fe/Mn incorporation. Due to the continuous decrease in oxygen vacancy for Fe and Mn-S0 samples, the spin-orbit splitting is also increasing continuously.

The $Bi^{5+}$ 4$f_{7/2}$ and $Bi^{5+}$ 4$f_{5/2}$ peak is observed at (161.96 eV, 165.51 eV) for Fe-S0, (162.85 eV and ~167.94 eV) for S0 and finally at (162.22 eV, 167.41 eV) for Mn-S0 [43,44]. The spin-orbit splitting energy for $Bi^{5+}$4$f_{7/2}$ and $Bi^{5+}$4$f_{5/2}$ was observed to be increased more from 5.10 eV for S0 to 5.31 eV for Fe-S0 and 5.19 eV for Mn-S0 samples. The decrease in B.E for Fe/Mn S0 samples is because the bonding of $Bi^{5+}$ with $O^{2-}$ is weakening. The presence of undesired $Bi^{5+}$ is there to compensate for the inherent A-site vacancy produced due to the high-temperature sintering of ceramic samples. Many works have indicated that these peaks could be due to the relaxed or defect states of Bismuth [45,46]. The area ratio of $f_{7/2}$ peak to $f_{5/2}$ peak for both $Bi^{3+}$ and $Bi^{5+}$ states is found to be ~4:3 for all the samples [40,43]. The fraction of $Bi^{5+}$ to $Bi^{3+}$ is highest for S0 (~0.78) and reduces considerably to 0.48 for Fe-S0 and drastically to 0.04 for Mn-S0. This may be a consequence of a decrease in oxygen vacancy in the Fe/Mn-S0 samples, which will be discussed below.

The Bi satellite peak detected in the Ti main peak region at 466.14 eV is observed to be continuously shifted to lower B.E of 466.10 eV Fe-S0 and 466.03 eV for Mn-S0 samples, respectively [38,47]. This is in good agreement with the decrease in oxygen vacancy in the respective samples. Such shifting of Bi-satellite peaks is affected due to the weakening of the $A_1$-O/ $A_4$-O bond strength in the respective samples.

iv.    Ba $3d$ core-level spectra:

The Ba $3d$ spectrum was deconvoluted to four peaks for all the samples [Fig. 4(a)-(c)]. Two peaks positioned at ~779.6 eV and ~794.70 eV for S0 represent the $3d_{5/2}$ and $3d_{3/2}$ states of $Ba^{2+}$ respectively [48,49]. This peak was observed to be at higher binding energies of (779.63 eV,795.00 eV) for Fe-S0 and (779.90 eV, 795.16eV) and Mn-S0. The spin-orbit splitting energy of $3d_{5/2}$ and $3d_{3/2}$ was found to be ~15.3 eV – 15.4 eV [50]. The area ratio of $3d_{5/2}$ to $3d_{3/2}$ was calculated to be ~1.5 for S0 and Mn-S0 while it is ~1.56 for the Fe-S0 sample [51]. The additional peaks which are present next to the main peaks represent the satellite peaks. The satellite peak positions increased from (780.73 eV, 796.10 eV) for S0 to (781.18 eV, 796.66 eV) for Fe-S0, and (782.28 eV, 797.57 eV) for Mn-S0 samples [49,52]. The ratio of satellite peak to main peak contribution is 0.69 for S0, while decreased to 0.49 for Fe-S0 and 0.45 for Mn-S0. Such consistent variation of binding energy and satellite peak contribution can be due to the following reason. Inherently, A-site vacancy is already there in the samples due to Na and Bi loss due to the high-temperature sintering of the ceramic samples. The shifting of binding energy to a higher value in the Fe/Mn samples signifies the presence of strong Ba-O bonds. This also reveals that the chance of Ba loss is decreasing which is also in good agreement with the reduction of the extra hump type region observed at ~791 eV [52,53]. Such behavior is also reflected in the oxygen vacancy variation for all the samples. The increase in the $A_2$-O bond strength influenced the $Ba^{2+}$ satellite peak to shift towards the higher binding energy peak position in the respective samples.

v.      Ti spectra:

The Ti ion is supposed to be in the $Ti^{4+}$ state. However, for the S0 sample, a presence of a mixed oxidation state of $Ti^{3+}$ and $Ti^{4+}$ was observed. A predominant $Ti^{4+}$ state was observed for the doped Fe-S0 and Mn-S0 samples [Fig. 4(d)-(f)]. For the S0 sample $Ti^{3+}$ $2p_{3/2}$ and $2p_{1/2}$ peaks are observed at 457.19 eV, 462.81 eV while, $Ti^{4+}2p_{3/2}$ and $Ti^{4+}$ $2p_{1/2}$ peaks are observed at (458.72 eV, 464.49 eV) [54,55]. The fraction of $Ti^{3+}$ to $Ti^{4+}$ fraction was found to be ~ 0.39 for S0. Hence, the presence of about 39% of Ti in the 3+ state instead of the desired 4+ state indicates the presence of associated oxygen vacancies. The Fe and Mn doping help to reduce the oxygen vacancy which is a bit counter-intuitive in this case as in general Fe and Mn are of a lesser valence state than Ti. However, a complete analysis of the valence states considering all the component ions indicates that such an observation is only possible when the A-site cations are lost due to the high-temperature annealing process, which Fe and Mn can compensate for.

The $Ti^{4+}2p_{3/2}$ and $Ti^{4+}$ $2p_{1/2}$ peak energies were observed to decrease to (457.90 eV, 463.68 eV) for Fe-S0 and (458.10 eV, 463.76 eV) for Mn-S0 samples. Hence, the decrease in binding energy was more for Fe-S0 than for the Mn-S0 sample. The spin-orbit splitting (5.77 eV for S0) is observed to increase slightly to 5.78 eV for Fe-S0 but decrease significantly to 5.66 eV for the Mn-S0 sample [55]. The number of electrons available in a $Fe^{2+/3+}$ state is higher electrons than the $Mn^{2+/3+}$ state. So, the extent of overlap of energy levels increases or decreases according to the availability of participating electrons. The spin-orbit splitting influences the optical band gap properties of a material. In general, the intensity of the peak is higher for the peak corresponding to a higher j value due to higher degeneracy. But, for the Ti spectra, a reverse phenomenon was observed for all the samples. Such behavior is due to the Coster- Kronig effect which is also explained as the special form of the Auger transition process in which the core-shell having a certain principal quantum number is quickly filled by an electron from a high energy shell of the same principal number [56–58]. For the present case, the area ratio of $Ti^{4+}$ $2p_{3/2}$ to $Ti^{4+}2p_{1/2}$ was calculated to be 0.57, 0.71, 1.13 for S0, Fe-S0, and Mn-S0 samples. The oxygen vacancy is observed to be decreased continuously for Fe-S0 and Mn-S0 samples. During the photoemission process, the oxygen vacancy influences the photoelectrons which are going to the continuum. In all the samples a peak is observed at ~468.47 eV which is called the plasmon peak of Ti [59]. Such peaks are generated due to the interactions of the emitted electrons with free electrons. These electrons are excited to the empty states that are known as collective oscillations, called plasmons [55,59]. The plasmonic peak contribution concerning Ti main peak contribution decreased from 0.41 for S0 to 0.21for Fe-S0 to 0.11 for Mn-S0. The decrease in plasmonic loss of Ti is due to the decrease in oxygen vacancy and decrease in the Coster-Kronig effect which is observed for all the samples.

The Ti-LMM Auger peaks observed with the Na-1$s$ main peak at 1070.59 eV for S0 tend to be significantly increased to 1071.04 eV for Fe-S0, and 1071.40 eV for Mn-S0 samples. The increase in binding energy corresponds to a decrease in kinetic energy. The binding energy of Ti $M_2$ $3p_{1/2}$ is ~32.6 eV, and for V $M_2$ $3p_{1/2}$ is ~37.2 eV. "Ti" exists in a mixed oxidation state of 3+ and 4+ in the S0 sample, while it is only in the 4+ state for the Fe and Mn samples. The Ti-LMM Auger transition involves the filling of the core hole ($L_3$ $2p_{3/2}$) by the $M_2$ $3p_{1/2}$ electron followed by the emission of electrons from $M_2$ $3p_{1/2}$. The no of electrons in the V $M_2$ $3p_{1/2}$ shell influences the Ti $M_2$ $3p_{1/2}$ shell which consequently results in the shifting of B.E. The oxidation state of V decreased for Fe-S0 and Mn-S0 continuously, which means a continuous increase in the electrons in the V $M_2$ $3p_{1/2}$ shell. This results in the shifting of Ti $M_2$ $3p_{1/2}$. The second LMM auger peak observed at 1074.26 eV for S0 tends to decrease slightly for Fe-S0 and increase slightly for the Mn-S0 sample. Such reverse behavior of shifting of this Auger binding energy for S0 and Fe-S0 can be due to the presence of $Ti^{3+}$ in the S0 sample.

The broad region in the spectra is due to the overlapping of $Bi^{3+}$ $4d_{3/2}$ peak (~465eV) with $Ti^{4+}$ $2p_{1/2}$ (~463.4 eV) peak, which is observed to be present in all the samples [47]. The Satellite peak corresponding to Bi was observed to be present at ~460.38 eV in all three samples which is discussed elaborately in the Bi-4$f$ core-level spectra section [60].

vi.     V 2$p$ core level spectra:

The V-2p XPS spectra were deconvoluted to four peaks for all the samples [Fig. 5(a)-(c)]. The XPS data shows the presence of a mixed oxidation state of V for all the samples. The binding energy corresponding to all the observed valence states is tabulated in Table III [61–63]. In S0 and Fe-S0 samples, the V shows a mixed oxidation state of $V^{5+}$ and $V^{4+}$. But, in the Mn-S0 sample, the V shows a mixed oxidation state of $V^{4+}$ and $V^{3+}$. The area ratio obtained for 2$p_{3/2}$ and 2$p_{1/2}$ states for all the samples was estimated to be ~2:1 [55,64] . The total charge of V decreased consistently from 4.82 to 4.07 to 3.82 for S0, Fe-S0, and Mn-S0 samples. The ionic radii of $V^{5+}$, $V^{4+}$, and $V^{3+}$ are 0.68Å, 0.72Å, and 0.78Å, respectively. The average ionic radii of Ti are greater than the average ionic radii of Fe and Mn. The decrease in the oxidation state implies that the V atom is trying to accommodate a larger space at Ti-site because of higher ionic radii of Ti in comparison to the Fe and Mn [65].

V. Mn 2$p$ core-level spectra:

The Mn-2$p$ XPS spectra was deconvoluted to five peaks associated with $2p_{3/2}$, $2p_{1/2}$ peaks of $Mn^{2+}$ (640.7eV, 651.92 eV) and $Mn^{3+}$ (642.43 eV, 653.64 eV) and one satellite peak at 647.2 eV corresponding to $Mn^{3+}$ [Fig. 6(a)] [34,38]. It was observed that there is a presence of mixed oxidation state of $Mn^{2+}$ and $Mn^{3+}$ in which the fraction of $Mn^{2+}$ to $Mn^{3+}$ is 0.70. The area ratio of $2p3/2$ to $2p1/2$ was calculated to be ~2:1. The spin-orbit splitting energy for $Mn^{2+}$ and $Mn^{3+}$ was found to be 11.52 eV, and 11.94 eV, respectively [66,67].

vii. Fe 2$p$ core-level spectra:

The Fe 2$p$ features are weak due to which the fitting became difficult. Fig.6(b) represents the obtained Fe 2$p$ core-level spectra. Hence, the peak positions and their origin is reported based on different works of literature available [68–70]. The Fe 2$p$ spectra revealed that it consists of a mixed oxidation state of $Fe^{2+}$ and $Fe^{3+}$.

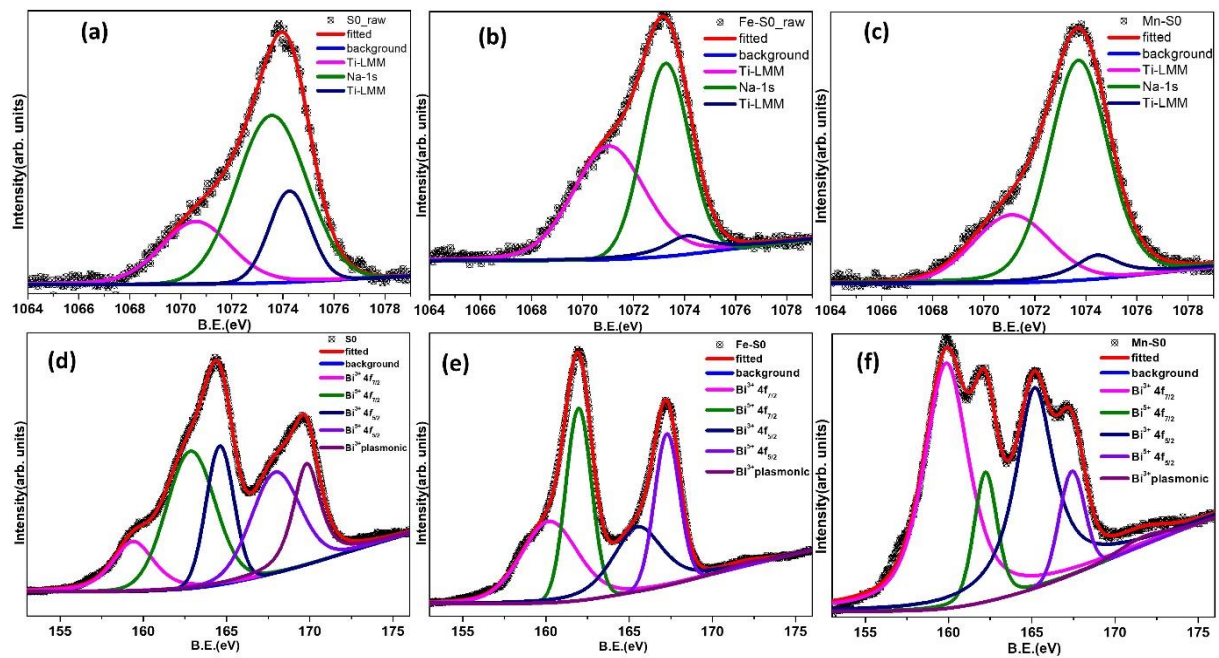

**Figure 3** Deconvoluted XPS core-level spectra of Na- 1$s$ for **(a)** S0 **(b)**Fe-S0 **(c)** Mn-S0 and Deconvoluted XPS core level spectra of Bi- 4$f$ for **(a)** S0 **(b)** Fe-S0 **(c)** Mn-S0 compositions

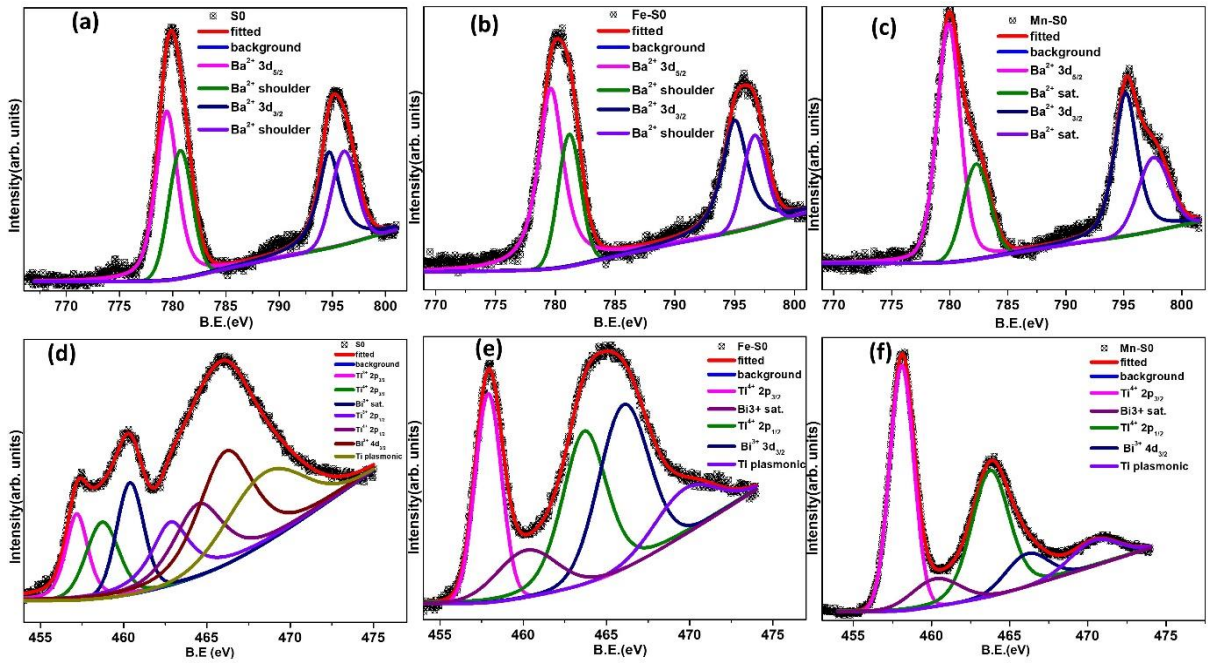

**Figure 4** Deconvoluted XPS core-level spectra of Ba $3d$ for **(a)** S0 **(b)** Fe-S0 **(c)** Mn-S0 and Deconvoluted XPS core level spectra of Ti $2p$ for **(a)** S0 **(b)** Fe-S0 **(c)** Mn-S0 compositions

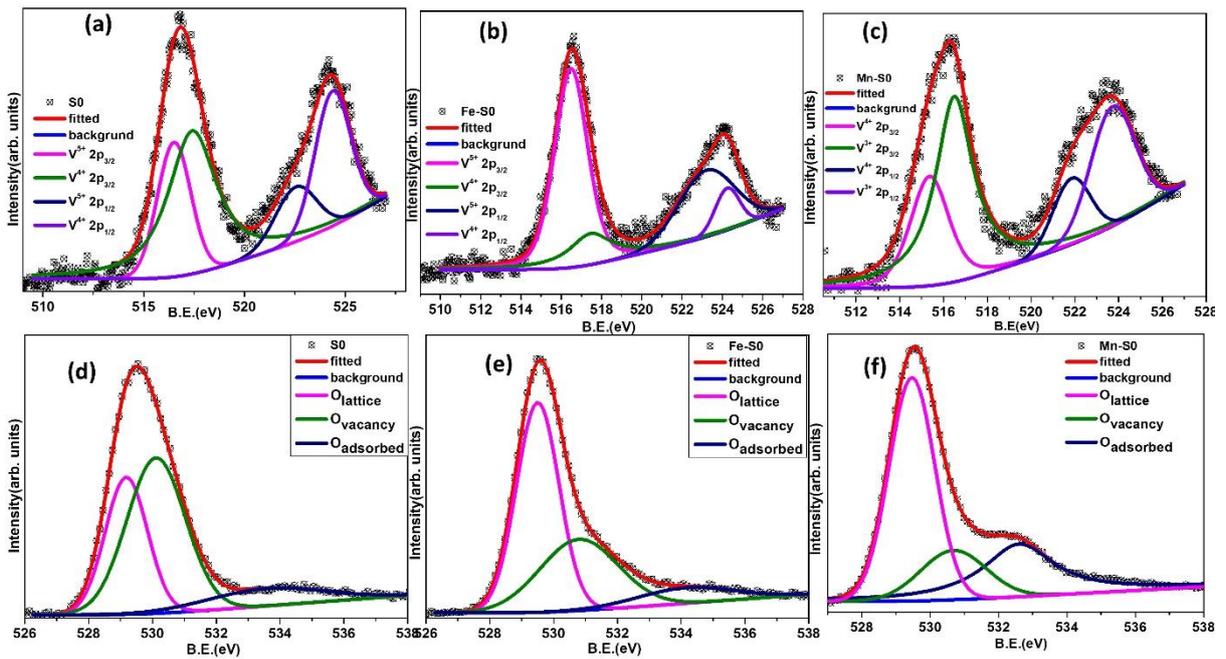

**Figure 5** Deconvoluted XPS core-level spectra of V $2p$ for **(a)** S0 **(b)** Fe-S0 **(c)** Mn-S0 samples and Deconvoluted XPS core-level spectra of O $1s$ for **(d)** S0 **(e)** Fe-S0 **(f)** Mn-S0 compositions

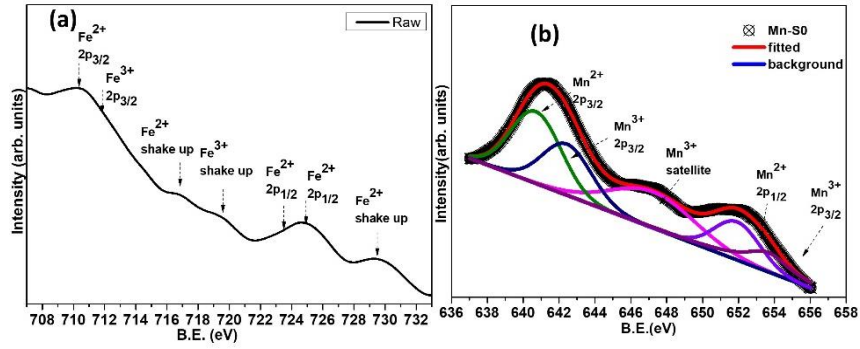

**Figure 6 (a)** XPS core-level spectra of Fe $2p$ for Fe-S0 sample and **(b)** Deconvoluted XPS core-level spectra of Mn $2p$ for Mn-S0

**Table III** Binding energy, FWHM of all peaks corresponding to each element present in all the three samples.

| Elements | Term states | S0 | | Fe-S0 | | Mn-S0 | |
|---|---|---|---|---|---|---|---|
| | | B.E(eV) | FWHM (eV) | B.E(eV) | FWHM (eV) | B.E(eV) | FWHM (eV) |
| Na | 1s | 1073.56 | 3.23 | 1073.26 | 3.32 | 1073.79 | 2.59 |
| Bi | $Bi^{3+}$ $4f_{7/2}$ | 159.40 | 2.95 | 160.26 | 3.19 | 159.86 | 3.21 |
| | $Bi^{5+}4f_{7/2}$ | 162.85 | 3.45 | 161.96 | 1.81 | 162.22 | 1.64 |
| | $Bi^{3+}$ $4f_{5/2}$ | 164.60 | 1.96 | 165.51 | 3.52 | 165.17 | 2.76 |
| | $Bi^{5+}4f_{5/2}$ | 167.94 | 3.47 | 167.27 | 1.77 | 167.41 | 1.80 |
| | Bi plasmonic | 169.81 | 1.93 | 171.96 | 2.3 | 171.69 | 3.19 |
| | $Bi^{3+}$ sat. | 460.38 | 1.93 | 460.20 | 2.50 | 460.27 | 3.63 |
| | $Bi^{3+}$ $4d_{3/2}$ | 464.14 | 3.40 | 465.55 | 3.54 | 466.10 | 3.20 |
| Ba | $Ba^{2+}$ $3d_{5/2}$ | 779.44 | 2.47 | 779.63 | 2.52 | 779.90 | 2.38 |
| | | 780.73 | 2.62 | 781.18 | 2.23 | 782.28 | 2.71 |
| | $Ba^{2+}3d_{3/2}$ | 794.70 | 2.48 | 795.00 | 2.56 | 795.16 | 2.37 |
| | | 796.10 | 2.79 | 796.66 | 2.31 | 797.57 | 3.16 |
| Ti | $Ti^{3+}$ $2p_{3/2}$ | 457.19 | 1.58 | - | - | - | - |
| | $Ti^{4+}2p_{3/2}$ | 458.72 | 2.31 | 457.90 | 1.87 | 458.10 | 1.78 |
| | $Ti^{3+}$ $2p_{1/2}$ | 462.49 | 2.80 | | | | |
| | $Ti^{4+}2p_{1/2}$ | 464.14 | 3.40 | 465.55 | 3.54 | 466.10 | 3.20 |
| | Ti-LMM | 1070.59 | 3.23 | 1071.04 | 2.16 | 1071.40 | 3.49 |
| | Ti-LMM | 1074.26 | 1.89 | 1074.06 | 2.08 | 1074.29 | 4.16 |

| | | | | | | |
|---|---|---|---|---|---|---|
| V | $V^{3+}$ $2p_{3/2}$ | - | - | - | - | 515.37 | 2.09 |
| | $V^{4+}$ $2p_{3/2}$ | 516.51 | 1.96 | 516.48 | 1.53 | 516.49 | 2.08 |
| | $V^{5+}2p_{3/2}$ | 517.41 | 2.87 | 517.45 | 1.75 | - | - |
| | $V^{3+}$ $2p_{1/2}$ | - | - | - | - | 521.86 | 2.02 |
| | $V^{4+}2p_{3/2}$ | 522.56 | 2.41 | 523.21 | 2.68 | 523.72 | 2.39 |
| | $V^{5+}2p_{1/2}$ | 524.37 | 2.08 | 524.30 | 1.42 | - | - |
| Mn | $Mn^{2+}2p_{3/2}$ | - | - | - | - | 640.70 | 3.30 |
| | $Mn^{3+}2p_{3/2}$ | - | - | - | - | 642.43 | 3.23 |
| | $Mn^{3+}$ sat. | - | - | - | - | 647.20 | 5.54 |
| | $Mn^{2+}2p_{1/2}$ | - | - | - | - | 651.92 | 3.00 |
| | $Mn^{3+}2p_{1/2}$ | - | - | - | - | 653.64 | 2.98 |
| Fe | $Fe^{2+}2p_{3/2}$ | - | - | 710.41 | - | - | - |
| | $Fe^{3+}2p_{3/2}$ | - | - | 711.94 | - | - | - |
| | $Fe^{3+}$ sat. | - | - | 716.80 | - | - | - |
| | $Fe^{3+}$ sat. | - | - | 719.72 | - | - | - |
| | $Fe^{2+}2p_{1/2}$ | - | - | 723.71 | - | - | - |
| | $Fe^{3+}2p_{1/2}$ | | | 725.23 | - | | |
| O | $O_L$ | 529.17 | 1.56 | 529.50 | 1.57 | 529.45 | 1.57 |
| | $O_V$ | 530.11 | 2.16 | 530.78 | 2.89 | 530.37 | 2.27 |
| | $O_A$ | 533.57 | 4.31 | 534.14 | 3.51 | 532.58 | 2.53 |

Dielectric Measurements:

To study different electrical properties of the prepared materials the capacitance (C), dielectric loss (tan δ), impedance (z), and phase angle were measured as a function of frequency (10Hz-1MHz) and temperature (50-450 ºC). The frequency-dependent dielectric constant at different temperatures is shown in Fig. 7(a)-(c). It was observed that the permittivity decreased with an increase in frequency. The space charge polarization at the grain boundaries is due to the presence of a potential barrier and is active at low frequencies [71]. Hence, the dielectric permittivity was observed to be highest at lower frequencies and decreased with an increase in frequency. It was also observed that the dielectric dispersion changes with temperature and becomes more evident at higher temperatures which is a characteristic behavior of ferroelectric materials [72]. The loss was also observed to be decreasing with an increase in frequency [Fig.7(d)-(f)]. The dielectric constant of the prepared samples increased according to the order of S0→Fe-S0→Mn-S0. This indicates that the Fe/Mn doping significantly improved the dielectric constant of the materials.

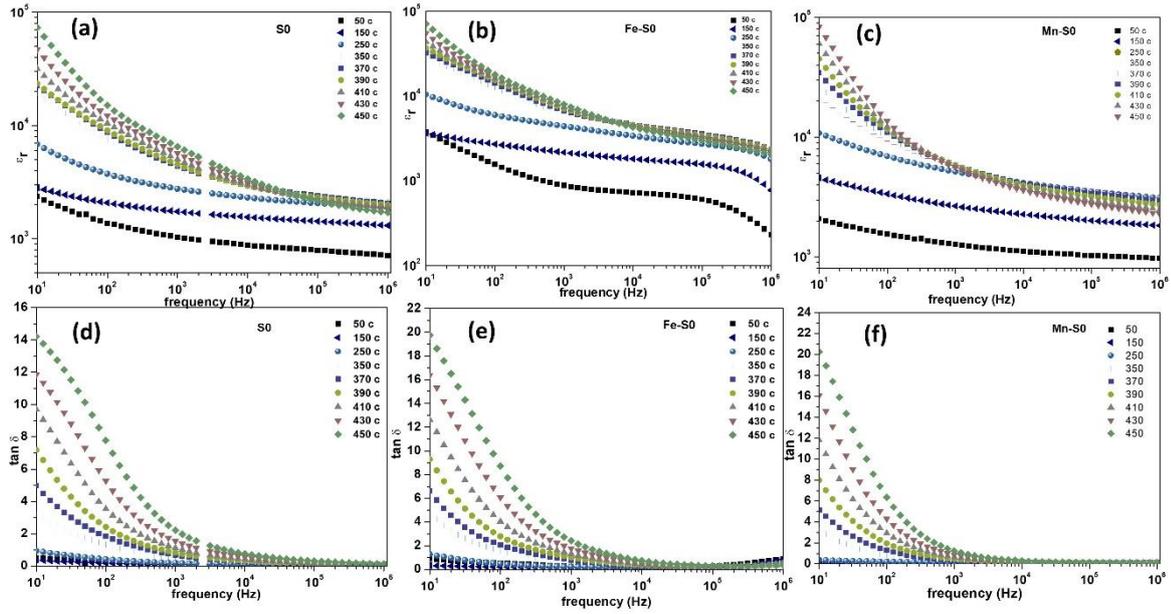

**Figure 7** Frequency dependent dielectric constant plot for **(a)** S0 **(b)** Fe-S0 **(c)** Mn-S0 and frequency dependent tan δ plots for **(d)** S0 **(e)** Fe-S0 **(f)** Mn-S0 compositions

Complex Impedance Spectroscopy Analysis:

The complex impedance spectroscopy is a technique to investigate the electrical microstructure–property relationship in different types of materials over a wide range of temperatures. Such studies are required to interpret whether the relaxation phenomena are due to long-range conductivity (delocalized) or dipole relaxation (localized) of materials [73]. The use of the imaginary part of the impedance ($Z''$) is useful for conductive analysis, when the long-range (delocalized) conductivity is dominant, whereas the imaginary part of the permittivity and electrical modulus $M''$ is suitable when localized relaxation dominates [74]. The electrical parameters obtained from the complex impedance technique can be expressed in terms of complex impedance ($Z^*$), complex admittance ($Y^*$), complex permittivity ($\varepsilon^*$), and complex electric modulus ($M^*$). The relations between these can be understood by the following equations [75]:

Complex impedance $Z^* = Z' - jZ''$

Complex admittance $Y^* = Y' + jY'' = \dfrac{1}{Z^*}$

Complex permittivity $\varepsilon^* = \varepsilon' - j\varepsilon'' = \dfrac{1}{j\omega C_0 Z^*}$

Complex Modulus $\left(\left(M^*\right) = M^{'} + jM^{''} = j\omega C_o Z^*\right.$

here, j= $\sqrt{-1}$ is the imaginary factor, $\omega = 2\pi f$ is the angular frequency and $C_0$ is a constant with dimensions of a capacitance.

Real ($Z'$) and Imaginary ($Z''$) part of Impedance study:

The frequency-dependent real part of modulus ($Z'$) [Fig. 8(a)-(c)] reveals that it is dispersive for all temperatures and the dispersive nature decreases at higher temperatures. The temperature dependence of $Z'$ decreases at a higher frequency and appears to merge. The $Z'$ the value decreased with an increase in temperature for all the compositions which implies NTCR type behavior [76]. The frequency-dependent imaginary part of the impedance ($Z''$) at different temperatures for all the compositions is shown in Fig. 8(d)-(f). A similar type of behavior was observed for all compositions. A peak appears in the temperature-dependent $Z''$ data for a particular frequency at and above a particular temperature, $T_{max}$ in the investigated frequency region(10Hz-1MHz) for each sample in the impedance spectra. The frequency-dependent $Z''$ data acquired at a particular temperature revealed information about the relaxation frequency, $f_{max}$. The $f_{max}$ is dependent on both composition and temperature. The $f_{max}$ is an intrinsic material property, and it does not depend on the sample geometrical factors [77]. With an increase in temperature $Z''$ decreases. The $f_{max}$ shifts to higher frequencies with temperature. This is a clear indication of a thermally activated electrical relaxation phenomenon [78]. A $Z''/Z''_{max}$ (normalized $Z''$ plot) is a better representation to understand the relaxation phenomena [Fig.9(a)-(c)]. This behavior is evidence of a small polaron hopping which is a consequence of an electron-lattice coupling. It is also observed that there is a broadening of peaks taking place with an increase in temperature. At higher frequencies, the $Z''$ values merge for all the temperatures which indicates the disappearance of space charge contribution. The frequency at which the peak appears called the relaxation time can be calculated from the relation $\omega_{max}\tau = 1$, where $\omega_{max} = 2\pi f_{max}$[74]. The $f_{max}$ shifts toward lower frequency and sharpens with Mn/Fe substitution. The Mn-S0 sample shows sharper and comparatively symmetric peaks than the Fe-S0 sample. A plot of temperature dependence of $f_{max}$ reveals an Arrhenius behaviour: $f_{max} = f_0 e^{\frac{-E_a}{k_\beta T}}$ [79].

Hence, the activation energy, $E_a$, was extracted from a linear fit of the logarithmic plot (i.e., $\ln[f_{max}]$ $vs$ $1/T$) of the above equation. The relaxation time, $\tau_0 = 1/2\pi f_0$, was also calculated from the fitting parameters. The relaxation time, $\tau_0$, decreased in the order $\tau_{0\text{-}Fe\text{-}S0}$ $\rightarrow \tau_{0\text{-}Mn\text{-}S0} \rightarrow \tau_{0\text{-}S0}$. The value of $\tau_{0\text{-}S0}$ is two orders smaller than $\tau_{0\text{-}Fe\text{-}S0}$ and $\tau_{0\text{-}Mn\text{-}S0}$. According to the *point defect relaxation theory*, oxygen vacancy, $V_O$, mobility, $\mu_O$, can be related to $\tau_0$. A shorter $\tau_0$ implies faster charge transport and hence, a greater possibility of movement of $V_O$ [80]. Hence, $\mu_O$ is higher for the undoped samples than the doped samples, with Fe-S0 having the lowest $\mu_O$ suggesting inhibition of oxygen migration by Fe/Mn doping.

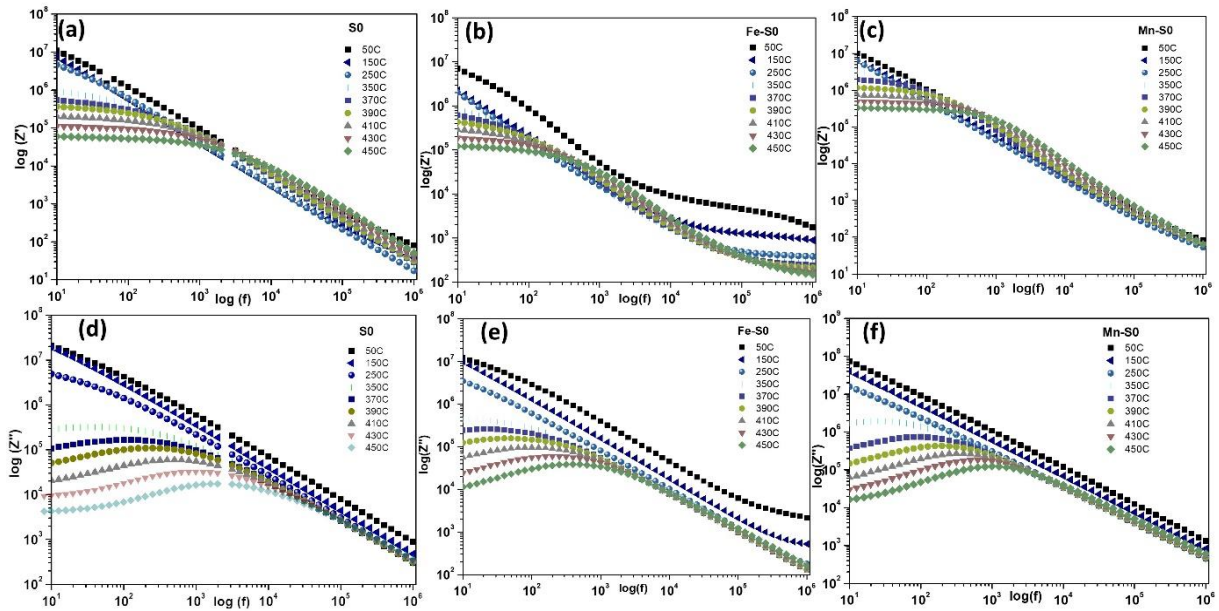

**Figure 8** Frequency-dependent Real part of impedance for **(a)** S0 **(b)** Fe-S0 **(c)** Mn-S0, and frequency-dependent imaginary part of impedance plots for **(d)** S0 **(e)** Fe-S0 **(f)** Mn-S0 compositions

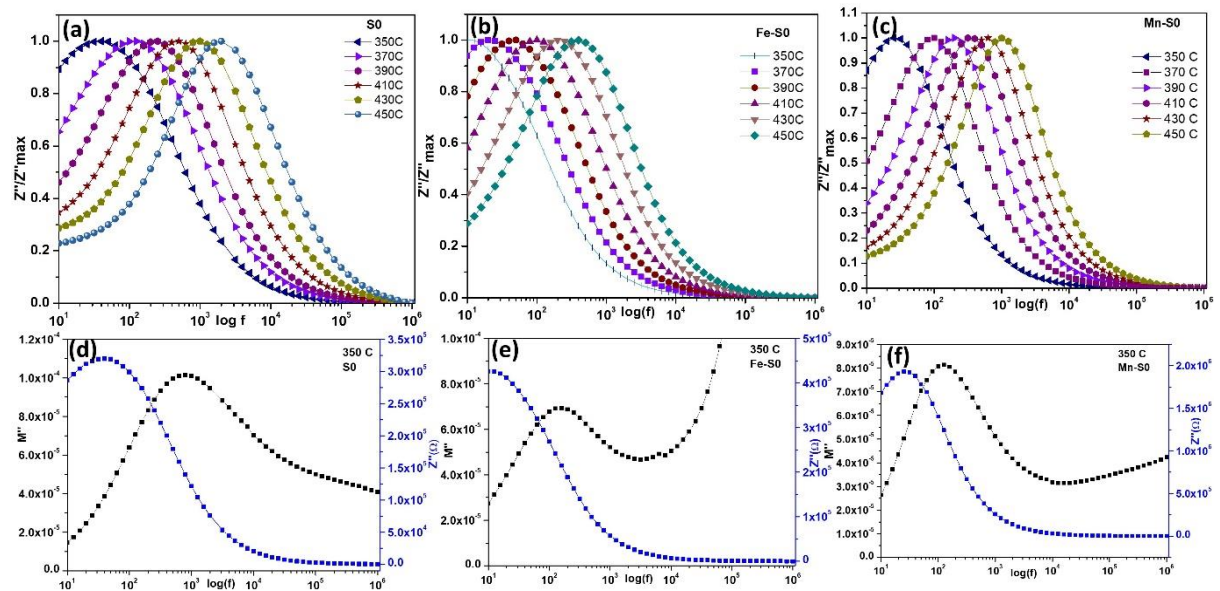

**Figure 9** Frequency dependent normalized Z″ plots at different temperatures (350°c-450°c) for **(a)** S0 **(b)** Fe-S0 **(c)** Mn-S0 and Merged plots of Z″ and M″ for **(d)** S0 **(e)** Fe-S0 **(f)** Mn-S0 compositions

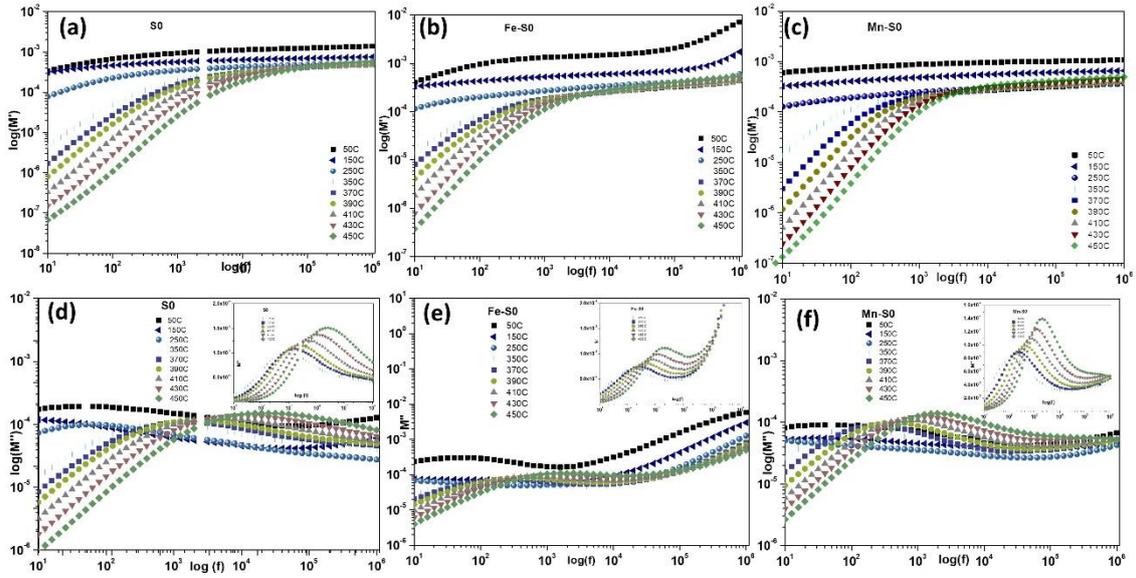

**Figure 10** Frequency dependent M' plots for **(a)** S0 **(b)** Fe-S0 **(c)** Mn-S0 and Frequency dependent M" plots for**(d)** S0 **(e)** Fe-S0 **(f)** Mn-S0 compositions

Real ($M'$) and Imaginary ($M''$) part of Modulus Spectra:

The real ($M'$) and imaginary parts ($M''$) of the modulus $M = M' + iM''$, was calculated from the impedance using the relations: $M' = \omega C_0 Z''$ and $M'' = \omega C_0 Z'$. The complex modulus data provides an insight on the dynamical behaviour of the materials by significantly suppressing the electrode polarization effect [81]. The frequency dependent$M'$[Fig. 10(a)-(c)] reveals low values at lower frequencies for all the three samples. This implies that the electrode effects are negligible at lower frequencies and can be ignored in the modulus formalism. $M'$increases linearly at very high frequencies for all temperatures. A linearly increasing trend of $M'$ indicates that the Z" value decreases proportionately which is evident from the Z" studies, implying that the dielectric loss that is the energy loss due to conduction is reduced. The occurrence of such a trend hint at a non-compatibility of the lattice to the frequency of the electric field applied, thereby reducing its response. The $M'$values decreased with an increase in temperature in the low-frequency region. At and above $f_{max}$ the $M'$ shows a reverse trend with the temperature that is the $M'$ increased with an increase in temperature. This agrees with the above analysis as thermal energy can assist the lattice in responding to the high-frequency electric field.

In Fig. 10(d)-(f) the imaginary part of the electric modulus is plotted as a function of frequency at different temperatures for all the compositions. The frequency-dependent $M''$plots reveal a relaxation feature in the form of a single peak at $f_{max}$ for all compositions only for temperatures greater than $T_{max}$ for the studied frequency region (10Hz-1MHz). The decreasing trends of $f_{max}$ with temperature and the decreasing trend of relaxation time, $\tau_0$ in the order, $\tau_{0-Mn-S0} \rightarrow \tau_{0-Fe-S0} \rightarrow \tau_{0-S0}$ with Mn/Fe substitution.

From the frequency-dependent $Z''$ and $M''$ plots, the lowest capacitance, and the largest resistance of the samples for a particular temperature can be identified. This plot enables one to differentiate the nature of transport properties, whether it belongs to a short-range (defect relaxation) or long-range (ionic/electronic conductivity) type of mobility of charge carriers, from the presence or absence of coincidence of peaks of the frequency-dependent $Z''$ and $M''$ plots, respectively [73,75,78]. For a long-range mobility process, the relaxation peaks in the frequency-dependent $M''$ and $Z''$ are coincident, while for a short range (localized) conduction process, these peaks are separated and are a non-Debye type relaxation process. A mismatch was observed between the relaxation peaks of the $Z''$ and $M''$ for all the compositions [Fig. 9(d)-(f)] with the $Z''$ relaxation peak at a low-frequency region while $M''$ peak at a high frequency region. The significant separation between the relaxation peaks implies the presence of localized movement of charge carriers and departure from ideal Debye-like behavior for all the samples.

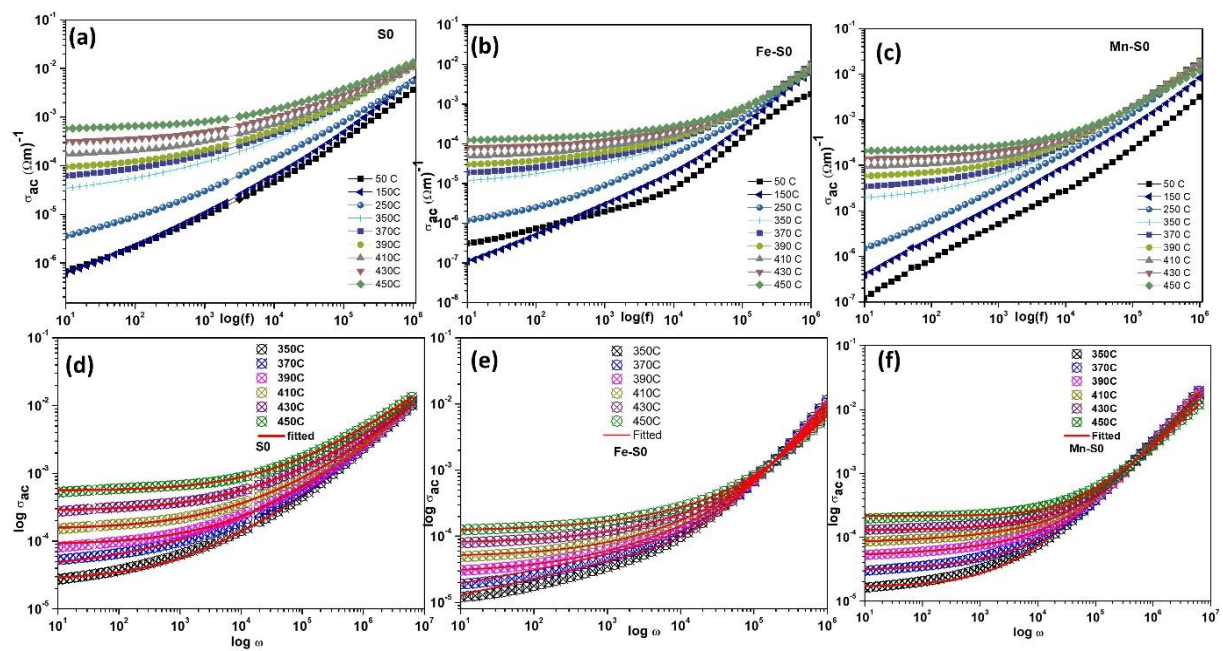

**Figure 11** Frequency-dependent AC conductivity for **(a)**S0 **(b)**Fe-S0 **(c)**Mn-S0, and fitted conductivity curves using JPL double power law for **(d)** S0 **(e)** Fe-S0 **(f)** Mn-S0 compositions

AC Conductivity ($\sigma_{ac}$) study:

The ac conductivity was estimated from the frequency-dependent dielectric transport properties of a material. The ac conductivity is calculated using the expression: $\sigma_{ac} = \omega\varepsilon_0\varepsilon_r tan\delta$, where, $\sigma_{ac}$ is the ac conductivity, $\varepsilon_0$, is the dielectric permittivity in a vacuum, $\varepsilon_r$, is the relative permittivity, and $\omega$ is the angular frequency. The ac conductivity ($\sigma_{ac}$) increased with an increase in frequency value showing dispersion throughout the frequency range at low-temperature regions [Fig.11(d)-(f)]. For high temperatures above $T_{max}$, the conductivity is moderately invariant in the low-frequency region below fmax. These values being close to the extremely low-frequency conductivity values correspond to the dc conductivity, $\sigma_{dc}$. However, at higher frequencies, the conductivity rises significantly (i.e., a dispersive behaviour) signifying a commendable contribution from the ac conductivity for all the compositions. In the S0 and Mn-S0 samples the conductivity seems to increase with frequency at extremely high frequencies, while for the Fe doped sample, the conductivity increases with frequency for frequencies higher than fmax, but seems to saturate, signifying a frequency-independent ac conductivity contribution at extremely high frequencies. The increase of $\sigma_{ac}$ with temperature implies a negative temperature coefficient of resistance (NTCR) behavior in all the samples [82].

To explain the conduction mechanism in detail, the frequency-dependent ac conductivity data is fitted with Jonscher's power law: $\sigma_{ac} = \sigma_{dc} + A\omega^n$, where, A is the temperature-dependent pre-exponential factor defining the strength of the polarizability and $0\leq n\leq 1$ is an exponent term providing the information about the degree of interaction between the mobile ions with the neighboring rigid lattice. These interactions can be either of translational type or localized in nature. The Jonscher's power law could not be fitted to the experimental data accurately for all the samples. The contribution of different types of phenomena at the microscopic domain may be the reason behind such a deviation from the general form of Jonscher's power law. The data could be fitted with a double power law: $\sigma_{ac} = \sigma_{dc} + A\omega^n + B\omega^m$, where, "n" ($0\leq n\leq 1$) and "m" ($1\leq m\leq 2$) represent the low and high-frequency regions. The plateau region at low frequencies is ascribed to the long-range translational motion of ions ($\sigma_{dc}$), whose origin is due to the successful hopping of the ions to the nearest available site due to the availability of long-time. On the other hand, the high-frequency conductivity spectra originated due to the two contending relaxation processes. In the first process due to the insufficient availability of time, the ions cannot relax, due to which the hopping is unsuccessful that is particularly due to the non-collinearity of the forward and backward motion of the ions, while in the second process successful hopping takes place when the ions are moved to a new site, and they get enough time to relax. So, such dispersive behavior in ac conductivity is a consequence of the increase in the ratio of successful to unsuccessful hoping rates [15,77,78].

The obtained dc conductivity values were further analyzed as a function of temperature to calculate the activation energy ($E_a$) using the Arrhenius equation as follows: $\sigma_{dc} = \sigma_0 e^{\frac{-E_a}{k_\beta T}}$. These values of $E_a$ obtained from Impedance, Modulus, and DC conductivity are tabulated in Table IV. The activation energy calculated from the impedance spectra is in the range of ~1.32-1.52, the modulus spectra 1.00-1.2 eV, while the conductivity spectra ~ 0.90-1.07 eV [Fig.12 (a)-(c)]. These ranges of $E_a$ values indicate highly resistive materials. Hence, the different $E_a$ values derived from $Z''$ and $M''$ hints at different species taking part in the conduction and relaxation process.

For perovskite materials, the ionic species are either singly ionized or doubly ionized oxygen vacancies ($V_O$s) which contribute to the relaxation and conduction process. For the singly ionized $V_O$s the $E_a$ lies in the range of 0.3- 0.5 eV, but for the doubly ionized $V_O$s, the activation energy is ~1 eV [83]. In these samples, the $E_a$ is ~ 1 eV, implying the presence of double-ionized $V_O$s. The ionization of $V_O$s generates conducting electrons which can be explained using the following relation:$V_o \leftrightarrow V_o^\bullet + e'$ and further $V_o^\bullet \leftrightarrow V_o^{\bullet\bullet} + e'$ [84]. The generated $e'$ might be responsible for the mixed oxidation state of $Ti^{4+}/Ti^{3+}$ and $V^{5+}/V^{4+}$ according to the following equations:$Ti^{4+} + e' \leftrightarrow Ti^{3+}$ and$V^{5+} + e' \leftrightarrow V^{4+}$ [78]. In the case of the Fe/Mn-doped sample, no mixed-oxidation state of $Ti^{4+}$ was observed. So, the $e'$ produced due to the oxygen vacancy could be the possible reason for $Fe^{2+}/Mn^{2+}$ and $Fe^{3+}/Mn^{2+}$ mixed oxidation state: $Fe^{3+} + e' \leftrightarrow Fe^{2+}$and $Mn^{3+} + e' \leftrightarrow Mn^{2+}$. The $V_o^{\bullet\bullet}$ are mobile in nature which increases the conductivity in a sample. But the presence of $Mn^{2+/3+}/Fe^{2+/3+}$ at the $Ti^{4+}$ site, is likely to promote the formation of $Mn_{Ti}''/Mn_{Ti}'$ and $Fe_{Ti}''/Fe_{Ti}'$ defects. These defects may attract the $V_o^{\bullet\bullet}$, thus creating $(Mn_{Ti}'' - V_o^{\bullet\bullet})/(Mn_{Ti}' - V_o^{\bullet\bullet})^\bullet$ and $Fe_{Ti}'' - V_o^{\bullet\bullet}/(Fe_{Ti}' - V_o^{\bullet\bullet})^\bullet$ defect dipoles which are less mobile [28,85]. The formation of such defect dipoles may suppress the mobility of $V_o^{\bullet\bullet}$, thus reducing the conductivity and improving the dielectric constant and relaxation time.

**Table IV** Activation energy calculated from dc conductivity plot, impedance spectra, and modulus spectra using Arrhenius relation.

| Composition | $E_a$ ($\sigma_{dc}$) | Z'' | | M'' | |
|---|---|---|---|---|---|
| | | $E_a$ (eV) | $\tau_0$ (s) | $E_a$ (eV) | $\tau_0$ (s) |
| S0 | 1.07 eV | 1.53 eV | $9.15 \times 10^{-15}$ | 1.27 eV | $1.01 \times 10^{-14}$ |
| Fe-S0 | 0.93 eV | 1.40 eV | $5.62 \times 10^{-14}$ | 1.00 eV | $8.12 \times 10^{-12}$ |
| Mn-S0 | 0.90 eV | 1.32 eV | $8.00 \times 10^{-14}$ | 1.05 eV | $3.24 \times 10^{-12}$ |

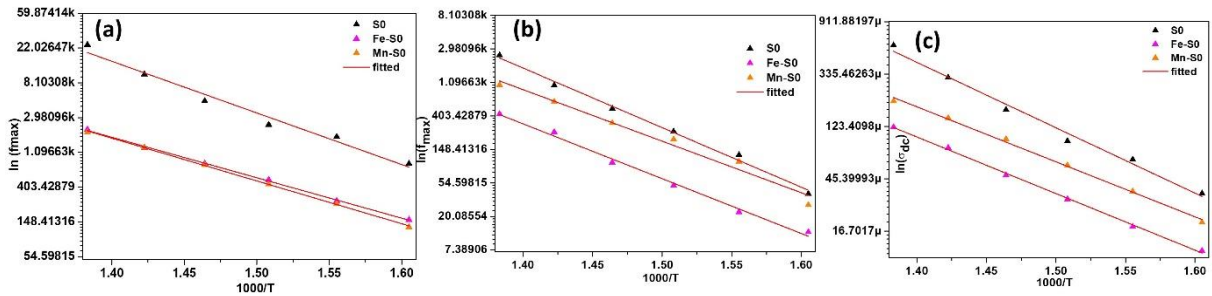

**Figure 12** Linear fit of ln($f_{max}$) and 1000/T for activation energy calculations from (a) Impedance spectroscopy and (b) Modulus spectra for all the compositions and (c) Linear fit of ln($\sigma_{dc}$) and 1000/T for activation energy calculations for all the compositions

Electrocaloric effect (ECE):

The electrocaloric effect is a physical phenomenon observed in materials that require the coupling of the electric field to dipolar order parameters. The change in electric field results in a change of the dipolar order thus changes of subsystem entropy [21]. The electrocaloric effect for all the samples was studied using an indirect method from the temperature-dependent PE loop [Fig. 13(a)-(c)]. Using the Maxwell relation: $(\frac{\partial S}{\partial E})_T = (\frac{\partial P}{\partial T})_E$ , the isothermal entropy change ($\Delta S$) and the adiabatic temperature change ($\Delta T$) due to electrocaloric effect can be calculated using the following relations [20,86]:

$$\Delta S = -\frac{1}{\rho} \int_{E_1}^{E_2} (\frac{\partial P}{\partial T})_E \, dE; \qquad \Delta T = -\frac{1}{\rho} \int_{E_1}^{E_2} \frac{T}{C_P} (\frac{\partial P}{\partial T})_E \, dE$$

where "$\rho$" is the density of the samples, "$C_p$" is the specific heat capacity (value taken from literature), "$\partial P/\partial T$" is the pyroelectric coefficient, and the external electric field ranges from $E_1$ to $E_2$ [18,26]. For the ECE calculation, the positive polarization and positive electric field values in the first quadrant were considered. where "$\rho$" is the density of the samples, "$C_p$ is the specific heat capacity (value taken from literature), "$\partial P/\partial T$" is the pyroelectric coefficient, and the external electric field ranges from $E_1$ to $E_2$. The polarization vs temperature plot is presented in Fig.14 (a)-(c) for all the samples. The polarization increased with an increase in the applied electric field. With an increase in the temperature, the polarization increased up to a certain value and further decreased. In fig. 15 (d)-(f), the first derivative of polarization with temperature shows that it increased up to a certain temperature and then decreased. Generally, NBT-6BT remains in the rhombohedral (*R3c*) phase at room temperature and with an increase in temperature it shows a mixed phase of tetragonal (*P4bm*) and rhombohedral (*R3c*), and then at further high temperature, it completely converts to tetragonal (*P4bm*). The direction of polarization in *R3c* is [111] while the direction of polarization in *P4bm* is [001]. Initially, the applied field is in the rhombohedral direction of polarization but as the temperature increases, the polarization direction is along [001] due to which non-collinearity of an applied field with the direction of polarization takes place which decreases the polarization. The negative electrocaloric effect was observed for all the samples [25,87,88]. The doped sample shows a higher negative electrocaloric effect as compared to the undoped one [Fig. 15]. The higher negative electrocaloric effect in the doped samples can be related to the defect dipoles. The $Fe^{2+}-V_O^{\bullet\bullet}$ / $Mn^{2+}-V_O^{\bullet\bullet}$ defect dipoles can produce a built-in electric field, which causes local order polarization regions around the defect dipoles. When the external electric field is applied, the local ordered regions become relatively disordered, which gives rise to the increase of isothermal changes in entropy ($\Delta S$) and the negative electrocaloric effect [85,89]. At high temperatures, the effect is approaching a positive electrocaloric effect which can be due to the associated phase transition involved in all the samples. This feature is common in all three samples. An additional feature was observed in the Fe-S0 sample around 310K, which can be due to some dipole rearrangement in the material. As the applied electric field increases, the change in entropy and change in temperature becomes more negative. The mixed positive/negative ECE is due to the associated phase transition in the NBT based samples. Table V shows the maximum ECE values obtained for all the samples at maximum applied electric field of 40 kV/cm. The Fe-S0 sample shows the highest temperature and entropy change followed by the Mn-S0 sample. The Mn/ Fe doped samples bear some potential for electrocaloric device application in refrigeration systems.

**Table V** Maximum $\Delta T$ and $\Delta S$ values obtained from indirect ECE calculation at maximum applied electric field.

| Materials | $\Delta T(K)$ | $\Delta S$ (J. $Kg^{-1}K^{-1}$) |
|---|---|---|
| $(Na_{0.5}Bi_{0.5})_{0.94}Ba_{0.06}Ti_{0.98}V_{0.02}O_3$ | -0.55 | -0.81 |
| $(Na_{0.5}Bi_{0.5})_{0.94}Ba_{0.06}Ti_{0.97}V_{0.02}Fe_{0.01}O_3$ | -1.16 | -1.65 |
| $(Na_{0.5}Bi_{0.5})_{0.94}Ba_{0.06}Ti_{0.98}V_{0.02}Mn_{0.01}O_3$ | -0.85 | -1.26 |

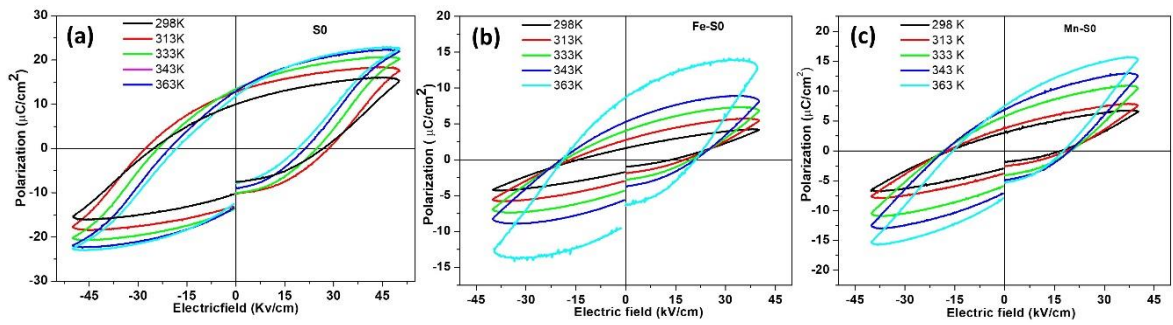

**Figure 13** Temperature dependent PE-loop for **(a)** V2 **(b)** V2-Fe1 **(c)** V2-Mn1

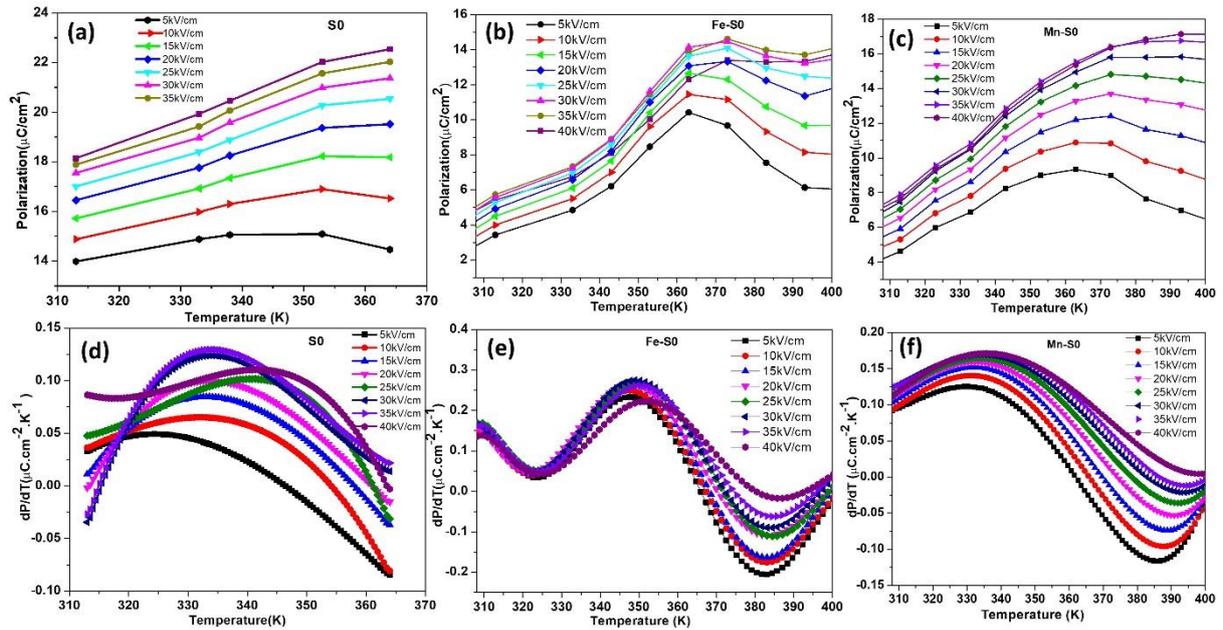

**Figure 14** Temperature-dependent polarization plots at the different applied electric field for **(a)**S0 **(b)**Fe-S0 **(c)**Mn-S0, and the temperature-dependent first derivative of polarization plots at the different applied electric field for **(d)** S0 **(e)** Fe-S0 **(f)** Mn-S0

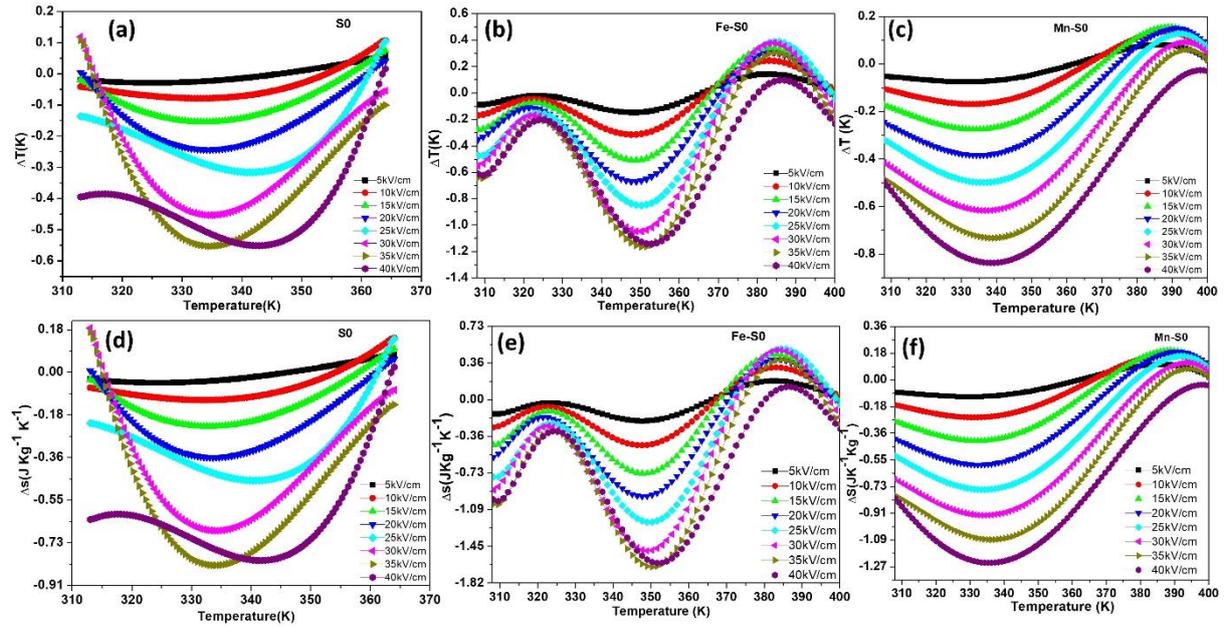

**Figure 15** Temperature-dependent adiabatic temperature change (T) plots at the different applied electric field for **(a)**S0 **(b)** Fe-S0 **(c)**Mn-S0, and temperature-dependent isothermal entropy change (S) at the different applied electric field for **(d)** S0 **(e)** Fe-S0 **(f)** Mn-S0

## CONCLUSIONS:

Rietveld refinement of XRD data of polycrystalline powders of $(Na_{0.5}Bi_{0.5})_{0.94}Ba0_{.06}Ti_{0.97}V_{0.02}M_{0.01}O_3$ (where, M = Ti, Fe, Mn) confirmed a *R3c* structure. The bandgap, Eg, reduced in the Fe-S0→S0→Mn-S0 order. This was a consequence of the changes in anti-phase octahedral tilt angle and the modifications in the spin-orbit splitting energy of the $Ti^{4+}2p_{3/2}$ and $Ti^{4+}2p_{1/2}$. The structural distortion derived from the Urbach energy continuously increased in the same order agreeing with the structural strain. The B.E and spin-orbit splitting energy for $Bi^{3+}4f_{7/2}$ and $Bi^{3+}4f_{5/2}$ followed an increasing trend in the order S0 → Fe-S0 → Mn-S0 samples implying a strengthening of the Bi-O bond. This is a consequence of the increased lattice oxygen content in the materials in the same order. $Bi^{5+}$ concentration in the order of S0 → Fe-S0 → Mn-S0 hints at a reduction of oxygen vacancy. The increased B.E. of Ba-3d feature in the Fe-S0/Mn-S0 samples signifies the presence of strong Ba-O bonds implying stronger $A_2$-O bonds. A mixed oxidation state of $Ti^{3+}$ and $Ti^{4+}$ was observed from the Ti-2p spectra of S0, while a predominant $Ti^{4+}$ state was observed for the doped Fe-S0 and Mn-S0 samples. This further strengthened the claim of lesser oxygen vacancy in the doped samples. The valence state of V decreased consistently from 4.82 to 4.07 to 3.82 in the order S0 → Fe-S0 → Mn-S0, as ionic size increases with higher valence state, the effective B-site size at the V-site increases with Fe/Mn doping and hence be able to compensate for the difference of ionic sizes between the host ion $Ti^{4+}$ and the V-ions. The Fe and Mn spectra showed the presence of a mixed state of $Mn^{2+}/Mn^{3+}$ and $Fe^{2+}/Fe^{3+}$ in the Mn-S0 and Fe-S0 samples, respectively. The dielectric constant ($\varepsilon_r$) significantly improved with doping for the Fe-S0 and Mn-S0 samples. The impedance and modulus spectra showed that the relaxation peak at and above $T_{max}$ revealed a prominent increase in the relaxation time ($\tau_0$) for the Fe-S0 and the Mn-S0 samples. The activation energy ($E_a$) calculated from the Modulus, Impedance, and the AC conductivity spectra revealed that the $E_a$ is ~ 1 eV, implying the presence of double-ionized $V_{OS}$. The different activation energy calculated from different spectra reveals that different charge species are taking part in the conduction and relaxation process. The mismatch in the $f_{ax}$ of $Z''$ and $M''$ suggests a non-Debye-like behavior and the dominance of short-range movement of defects. The electrocaloric effect (ECE) also revealed that there is a significant enhancement of the $\Delta S$ and the $\Delta T$ in both the Fe-S0 and Mn-S0 samples. The highest value of $\Delta T$ ~ -1.16 K and $\Delta S$ ~ -1.65 JK$^{-1}$Kg$^{-1}$ was obtained for the Fe-S0 sample. Such improvement of dielectric and ECE properties could be due to the formation of $(Mn''_{Ti} - V_o^{\bullet\bullet})/(Mn'_{Ti} - V_o^{\bullet\bullet})^{\bullet}$ and $Fe''_{Ti} - V_o^{\bullet\bullet}/(Fe'_{Ti} - V_o^{\bullet\bullet})^{\bullet}$ defect dipoles which are immobile/less mobile that is helping to suppress the mobility of $V_o^{\bullet\bullet}$, thus reducing the conductivity and improving the electrical properties.

SUPPLEMENTARY MATERIAL:

See supplementary material for the details of sample preparation and optical band gap study plots.

DECLARATION FOR CONFLICT OF INTEREST:

The authors declare no conflict of interest.

AUTHOR'S CONTRIBUTION:

K. S. Samantaray: Conceptualisation, Draft Writing, Analysis, Visualisation

R. Amin: Data Collection, Visualisation

E.G. Rini: Formal Analysis

I. Bhaumik: Data Collection, Supervision

A. Mekki: Data Collection, Analysis

K. Harrabi: Review

S. Sen: Supervision, Project administration, Review and editing

ACKNOWLEDGEMENT:

The authors would like to acknowledge the facilities at IIT Indore. A. Mekki, K. Harrabi, and S. Sen gratefully acknowledge the support of the King Fahd University of Petroleum and Minerals, Saudi Arabia, under the DF191055 DSR project. Ms. Koyal Suman Samantaray acknowledges MHRD for providing Prime Minister Research Fellowship (PMRF). Mr. Ruhul Amin acknowledges DST for INSPIRE Fellowship (No. IF160339). E.G Rini also acknowledges the DST, India for financial support under Women Scientist Scheme (SR/WOS-A/PM-99/2016 (G)).

AVAILABILITY OF DATA:

The data that support the findings of this study are available on request from the corresponding author. The data are not publicly available due to privacy and ethical reasons.

REFERENCES:

[1]    E. Aksel, J.L. Jones, Advances in Lead-Free Piezoelectric Materials for Sensors and Actuators, Sensors. 10 (2010) 1935–1954. https://doi.org/10.3390/s100301935.
[2]    S.K. Dey, R. Zuleeg, Integrated sol-gel PZT thin-films on Pt, Si, and GaAs for non-volatile memory applications, Ferroelectrics. 108 (1990) 37–46. https://doi.org/10.1080/00150199008018730.
[3]    S. Zhang, R. Xia, T.R. Shrout, Lead-free piezoelectric ceramics vs. PZT?, J Electroceram. 19 (2007) 251–257. https://doi.org/10.1007/s10832-007-9056-z.
[4]    T.R. Shrout, S.J. Zhang, Lead-free piezoelectric ceramics: Alternatives for PZT?, J Electroceram. 19 (2007) 113–126. https://doi.org/10.1007/s10832-007-9047-0.
[5]    L. Zhang, Y. Pu, M. Chen, T. Wei, X. Peng, Novel Na0.5Bi0.5TiO3 based, lead-free energy storage ceramics with high power and energy density and excellent high-temperature stability, Chemical Engineering Journal. 383 (2020) 123154. https://doi.org/10.1016/j.cej.2019.123154.
[6]    W.P. Cao, W.L. Li, X.F. Dai, T.D. Zhang, J. Sheng, Y.F. Hou, W.D. Fei, Large electrocaloric response and high energy-storage properties over a broad temperature range in lead-free NBT-ST ceramics, Journal of the European Ceramic Society. 36 (2016) 593–600. https://doi.org/10.1016/j.jeurceramsoc.2015.10.019.
[7]    A. Verma, A.K. Yadav, S. Kumar, V. Srihari, R. Jangir, H.K. Poswal, S. Biring, S. Sen, Enhanced energy storage properties in A-site substituted Na0.5Bi0.5TiO3 ceramics, Journal of Alloys and Compounds. 792 (2019) 95–107. https://doi.org/10.1016/j.jallcom.2019.03.304.
[8]    S. Swain, S. Kumar Kar, P. Kumar, Dielectric, optical, piezoelectric and ferroelectric studies of NBT–BT ceramics near MPB, Ceramics International. 41 (2015) 10710–10717. https://doi.org/10.1016/j.ceramint.2015.05.005.
[9]    S. Swain, P. Kumar, D.K. Agrawal, Sonia, Dielectric and ferroelectric study of KNN modified NBT ceramics synthesized by microwave processing technique, Ceramics International. 39 (2013) 3205–3210. https://doi.org/10.1016/j.ceramint.2012.10.005.

[10] A. Verma, A.K. Yadav, S. Kumar, V. Srihari, R. Jangir, H.K. Poswal, S. Biring, S. Sen, Structural, thermally stable dielectric, and energy storage properties of lead-free (1 − x)(Na0.50Bi0.50)TiO3 − xKSbO3 ceramics, J Mater Sci: Mater Electron. 30 (2019) 15005–15017. https://doi.org/10.1007/s10854-019-01873-1.

[11] A. Verma, A.K. Yadav, S. Kumar, V. Srihari, R. Jangir, H.K. Poswal, S.-W. Liu, S. Biring, S. Sen, Improvement of energy storage properties with the reduction of depolarization temperature in lead-free (1 − x)Na $_{0.5}$ Bi $_{0.5}$ TiO $_3$ - x AgTaO $_3$ ceramics, Journal of Applied Physics. 125 (2019) 054101. https://doi.org/10.1063/1.5075719.

[12] E. Aksel, H. Foronda, K.A. Calhoun, J.L. Jones, S. Schaab, T. Granzow, PROCESSING AND PROPERTIES OF Na $_{0.5}$ Bi $_{0.5}$ TiO $_3$ PIEZOELECTRIC CERAMICS MODIFIED WITH La , Mn AND Fe, Funct. Mater. Lett. 03 (2010) 45–48. https://doi.org/10.1142/S1793604710000877.

[13] J. Bubesh Babu, M. He, D.F. Zhang, X.L. Chen, R. Dhanasekaran, Enhancement of ferroelectric properties of Na1⁄2Bi1⁄2TiO3-BaTiO3 single crystals by Ce dopings, Appl. Phys. Lett. 90 (2007) 102901. https://doi.org/10.1063/1.2709917.

[14] Š. Svirskas, M. Ivanov, Š. Bagdzevičius, M. Dunce, M. Antonova, E. Birks, A. Sternberg, A. Brilingas, J. Banys, Dynamics of Phase Transition in 0.4NBT-0.4ST-0.2PT Solid Solution, Integrated Ferroelectrics. 134 (2012) 81–87. https://doi.org/10.1080/10584587.2012.665300.

[15] S.R. Kanuru, K. Baskar, R. Dhanasekaran, Synthesis, structural, morphological and electrical properties of NBT–BT ceramics for piezoelectric applications, Ceramics International. 42 (2016) 6054–6064. https://doi.org/10.1016/j.ceramint.2015.12.162.

[16] X. Liu, H. Yang, F. Yan, Y. Qin, Y. Lin, T. Wang, Enhanced energy storage properties of BaTiO3-Bi0.5Na0.5TiO3 lead-free ceramics modified by SrY0.5Nb0.5O3, Journal of Alloys and Compounds. 778 (2019) 97–104. https://doi.org/10.1016/j.jallcom.2018.11.106.

[17] H.S. Mohanty, T. Dam, H. Borkar, D.K. Pradhan, K.K. Mishra, A. Kumar, B. Sahoo, P.K. Kulriya, C. Cazorla, J.F. Scott, D.K. Pradhan, Structural transformations and physical properties of (1 − x ) Na $_{0.5}$ Bi $_{0.5}$ TiO $_3$ - x BaTiO $_3$ solid solutions near a morphotropic phase boundary, J. Phys.: Condens. Matter. 31 (2019) 075401. https://doi.org/10.1088/1361-648X/aaf405.

[18] Y. Bai, G.-P. Zheng, S.-Q. Shi, Abnormal electrocaloric effect of Na0.5Bi0.5TiO3–BaTiO3 lead-free ferroelectric ceramics above room temperature, Materials Research Bulletin. 46 (2011) 1866–1869. https://doi.org/10.1016/j.materresbull.2011.07.038.

[19] A. Greco, C. Masselli, Electrocaloric Cooling: A Review of the Thermodynamic Cycles, Materials, Models, and Devices, Magnetochemistry. 6 (2020) 67. https://doi.org/10.3390/magnetochemistry6040067.

[20] E. Birks, M. Dunce, J. Peräntie, J. Hagberg, A. Sternberg, Direct and indirect determination of electrocaloric effect in Na $_{0.5}$ Bi $_{0.5}$ TiO $_3$, Journal of Applied Physics. 121 (2017) 224102. https://doi.org/10.1063/1.4985067.

[21] Z. Kutnjak, B. Rožič, R. Pirc, Electrocaloric Effect: Theory, Measurements, and Applications, in: Wiley Encyclopedia of Electrical and Electronics Engineering, John Wiley & Sons, Inc., Hoboken, NJ, USA, 2015: pp. 1–19. https://doi.org/10.1002/047134608X.W8244.

[22] A. Mischenko, Q. Zhang, J.F. Scott, R.W. Whatmore, N.D. Mathur, Giant electrocaloric effect in thin film Pb Zr_0.95 Ti_0.05 O_3, Science. 311 (2006) 1270–1271. https://doi.org/10.1126/science.1123811.

[23] O. Turki, A. Slimani, L. Seveyrat, G. Sebald, V. Perrin, Z. Sassi, H. Khemakhem, L. Lebrun, Structural, dielectric, ferroelectric, and electrocaloric properties of 2% Gd $_2$ O $_3$ doping (Na $_{0.5}$ Bi $_{0.5}$ ) $_{0.94}$ Ba $_{0.06}$ TiO $_3$ ceramics, Journal of Applied Physics. 120 (2016) 054102. https://doi.org/10.1063/1.4960141.

[24] Y. Zhang, W. Li, Z. Wang, Y. Qiao, H. Xia, R. Song, Y. Zhao, W. Fei, Perovskite Sr $_{1-x}$ (Na $_{0.5}$ Bi $_{0.5}$ ) $_x$ Ti $_{0.99}$ Mn $_{0.01}$ O $_3$ Thin Films with Defect Dipoles for High Energy-Storage and Electrocaloric Performance, ACS Appl. Mater. Interfaces. 11 (2019) 37947–37954. https://doi.org/10.1021/acsami.9b14815.

[25] I. Bhaumik, S. Ganesamoorthy, R. Bhatt, A.K. Karnal, P.K. Gupta, S. Takekawa, K. Kitamura, Bipolar electro-caloric effect in $Sr_xBa_{(1-x)}Nb_2O_6$ lead-free ferroelectric single crystal, EPL. 107 (2014) 47001. https://doi.org/10.1209/0295-5075/107/47001.

[26] W.P. Cao, W.L. Li, D. Xu, Y.F. Hou, W. Wang, W.D. Fei, Enhanced electrocaloric effect in lead-free NBT-based ceramics, Ceramics International. 40 (2014) 9273–9278. https://doi.org/10.1016/j.ceramint.2014.01.149.

[27] T. Zhang, W. Li, Y. Hou, Y. Yu, W. Cao, Y. Feng, W. Fei, Positive/negative electrocaloric effect induced by defect dipoles in PZT ferroelectric bilayer thin films, RSC Adv. 6 (2016) 71934–71939. https://doi.org/10.1039/C6RA14776C.

[28] Z.-H. Zhao, Y. Dai, F. Huang, The formation and effect of defect dipoles in lead-free piezoelectric ceramics: A review, Sustainable Materials and Technologies. 20 (2019) e00092. https://doi.org/10.1016/j.susmat.2019.e00092.

[29] K.S. Samantaray, R. Amin, E.G. Rini, S. Sen, Fe-doped $Na0.47Bi0.47Ba0.06Ti0.98-xV0.02FexO3$: Structure correlated vibrational, optical and electrical properties, Journal of Alloys and Compounds. 848 (2020) 156503. https://doi.org/10.1016/j.jallcom.2020.156503.

[30] H.D. Megaw, C.N.W. Darlington, Geometrical and structural relations in the rhombohedral perovskites, Acta Cryst A. 31 (1975) 161–173. https://doi.org/10.1107/S0567739475000332.

[31] B.D. Cullity, Elements of x-ray diffraction, 2d ed, Addison-Wesley Pub. Co, Reading, Mass, 1978.

[32] H. He, X. Lin, S. Li, Z. Wu, J. Gao, J. Wu, W. Wen, D. Ye, M. Fu, The key surface species and oxygen vacancies in MnOx(0.4)-CeO2 toward repeated soot oxidation, Applied Catalysis B: Environmental. 223 (2018) 134–142. https://doi.org/10.1016/j.apcatb.2017.08.084.

[33] Q. Li, L. Yin, Z. Li, X. Wang, Y. Qi, J. Ma, Copper Doped Hollow Structured Manganese Oxide Mesocrystals with Controlled Phase Structure and Morphology as Anode Materials for Lithium Ion Battery with Improved Electrochemical Performance, ACS Appl. Mater. Interfaces. 5 (2013) 10975–10984. https://doi.org/10.1021/am403215j.

[34] E. Beyreuther, S. Grafström, L.M. Eng, C. Thiele, K. Dörr, XPS investigation of Mn valence in lanthanum manganite thin films under variation of oxygen content, Phys. Rev. B. 73 (2006) 155425. https://doi.org/10.1103/PhysRevB.73.155425.

[35] X. Dai, J. Cheng, Z. Li, M. Liu, Y. Ma, X. Zhang, Reduction kinetics of lanthanum ferrite perovskite for the production of synthesis gas by chemical-looping methane reforming, Chemical Engineering Science. 153 (2016) 236–245. https://doi.org/10.1016/j.ces.2016.07.011.

[36] F. Yang, M. Li, L. Li, P. Wu, E. Pradal-Velázquez, D.C. Sinclair, Defect chemistry and electrical properties of sodium bismuth titanate perovskite, J. Mater. Chem. A. 6 (2018) 5243–5254. https://doi.org/10.1039/C7TA09245H.

[37] C.D. Wagner, Auger lines in x-ray photoelectron spectrometry, Anal. Chem. 44 (1972) 967–973. https://doi.org/10.1021/ac60314a015.

[38] H. Zhang, H. Deng, C. Chen, L. Li, D. Lin, X. Li, X. Zhao, H. Luo, J. Yan, Chemical nature of giant strain in Mn-doped 0.94(Na0.5Bi0.5)TiO3–0.06BaTiO3 lead-free ferroelectric single crystals, Scripta Materialia. 75 (2014) 50–53. https://doi.org/10.1016/j.scriptamat.2013.11.017.

[39] P. Singh, P.K. Jha, A.S.K. Sinha, P.A. Jha, P. Singh, Ion dynamics of non-stoichiometric Na0.5+xBi0.5-xTiO3-δ: A degradation study, Solid State Ionics. 345 (2020) 115158. https://doi.org/10.1016/j.ssi.2019.115158.

[40] V.S. Dharmadhikari, S.R. Sainkar, S. Badrinarayan, A. Goswami, Characterisation of thin films of bismuth oxide by X-ray photoelectron spectroscopy, Journal of Electron Spectroscopy and Related Phenomena. 25 (1982) 181–189. https://doi.org/10.1016/0368-2048(82)85016-0.

[41] X. Li, Y. Sun, T. Xiong, G. Jiang, Y. Zhang, Z. Wu, F. Dong, Activation of amorphous bismuth oxide via plasmonic Bi metal for efficient visible-light photocatalysis, Journal of Catalysis. 352 (2017) 102–112. https://doi.org/10.1016/j.jcat.2017.04.025.

[42] A.K.R. Police, S.V.P. Vattikuti, K.K. Mandari, M. Chennaiahgari, P.S. M.V., D.K. Valluri, C. Byon, Bismuth oxide cocatalyst and copper oxide sensitizer in Cu2O/TiO2/Bi2O3 ternary

photocatalyst for efficient hydrogen production under solar light irradiation, Ceramics International. 44 (2018) 11783–11791. https://doi.org/10.1016/j.ceramint.2018.03.262.

[43] R. Zalecki, W.M. Woch, M. Kowalik, A. Kołodziejczyk, G. Gritzner, Bismuth Valence in a Tl $_{0.7}$ Bi $_{0.3}$ Sr $_{1.6}$ Ba $_{0.4}$ CaCu $_2$ O $_y$ Superconductor from X-Ray Photoemission Spectroscopy, Acta Phys. Pol. A. 118 (2010) 393–395. https://doi.org/10.12693/APhysPolA.118.393.

[44] K. Nakano, K. Oka, T. Watanuki, M. Mizumaki, A. Machida, A. Agui, H. Kim, J. Komiyama, T. Mizokawa, T. Nishikubo, Y. Hattori, S. Ueda, Y. Sakai, M. Azuma, Glassy Distribution of Bi $^{3+}$ /Bi $^{5+}$ in Bi $_{1−x}$ Pb $_x$ NiO $_3$ and Negative Thermal Expansion Induced by Intermetallic Charge Transfer, Chem. Mater. 28 (2016) 6062–6067. https://doi.org/10.1021/acs.chemmater.6b01160.

[45] S. Chaturvedi, I. Sarkar, M.M. Shirolkar, U.-S. Jeng, Y.-Q. Yeh, R. Rajendra, N. Ballav, S. Kulkarni, Probing bismuth ferrite nanoparticles by hard x-ray photoemission: Anomalous occurrence of metallic bismuth, Appl. Phys. Lett. 105 (2014) 102910. https://doi.org/10.1063/1.4895672.

[46] P.S.V. Mocherla, D. Prabhu, M.B. Sahana, N.Y. Hebalkar, R. Gopalan, M.S. Ramachandra Rao, C. Sudakar, High temperature magnetic studies on Bi $_{1-x}$ Ca $_x$ Fe $_{1-y}$ Ti $_y$ O $_{3-δ}$ nanoparticles: Observation of Hopkinson-like effect above T $_N$, Journal of Applied Physics. 124 (2018) 073904. https://doi.org/10.1063/1.5038007.

[47] J.R. D. E., R.A.U. Rahman, S. B., M. Ramaswamy, Room temperature multiferroicity and magnetoelectric coupling in Na-deficient sodium bismuth titanate, Appl. Phys. Lett. 114 (2019) 062902. https://doi.org/10.1063/1.5078575.

[48] S. Zhang, B. Zhang, S. Li, Z. Huang, C. Yang, H. Wang, Enhanced photocatalytic activity in Ag-nanoparticle-dispersed BaTiO3 composite thin films: Role of charge transfer, J Adv Ceram. 6 (2017) 1–10. https://doi.org/10.1007/s40145-016-0209-x.

[49] V. Craciun, R.K. Singh, Characteristics of the surface layer of barium strontium titanate thin films deposited by laser ablation, Appl. Phys. Lett. 76 (2000) 1932–1934. https://doi.org/10.1063/1.126216.

[50] B.C. Keswani, R.S. Devan, R.C. Kambale, A.R. James, S. Manandhar, Y.D. Kolekar, C.V. Ramana, Correlation between structural, magnetic and ferroelectric properties of Fe-doped (Ba-Ca)TiO3 lead-free piezoelectric, Journal of Alloys and Compounds. 712 (2017) 320–333. https://doi.org/10.1016/j.jallcom.2017.03.301.

[51] A. Rodrigues, S. Bauer, T. Baumbach, Effect of post-annealing on the chemical state and crystalline structure of PLD Ba0.5Sr0.5TiO3 films analyzed by combined synchrotron X-ray diffraction and X-ray photoelectron spectroscopy, Ceramics International. 44 (2018) 16017–16024. https://doi.org/10.1016/j.ceramint.2018.06.018.

[52] J.D. Baniecki, M. Ishii, T. Shioga, K. Kurihara, S. Miyahara, Surface core-level shifts of strontium observed in photoemission of barium strontium titanate thin films, Appl. Phys. Lett. 89 (2006) 162908. https://doi.org/10.1063/1.2357880.

[53] Y. Fujisaki, Y. Shimamoto, Y. Matsui, Analysis of Decomposed Layer Appearing on the Surface of Barium Strontium Titanate, Jpn. J. Appl. Phys. 38 (1999) L52–L54. https://doi.org/10.1143/JJAP.38.L52.

[54] L. de, SK. SEN*, I. RIGA ad J. VERBIST, (1976) 5.

[55] M.C. Biesinger, L.W.M. Lau, A.R. Gerson, R.St.C. Smart, Resolving surface chemical states in XPS analysis of first row transition metals, oxides and hydroxides: Sc, Ti, V, Cu and Zn, Applied Surface Science. 257 (2010) 887–898. https://doi.org/10.1016/j.apsusc.2010.07.086.

[56] D. Kim, C. Tóth, C.P.J. Barty, Population inversion between atomic inner-shell vacancy states created by electron-impact ionization and Coster-Kronig decay, Phys. Rev. A. 59 (1999) R4129–R4132. https://doi.org/10.1103/PhysRevA.59.R4129.

[57] P. Le Fèvre, J. Danger, H. Magnan, D. Chandesris, J. Jupille, S. Bourgeois, M.-A. Arrio, R. Gotter, A. Verdini, A. Morgante, Stoichiometry-related Auger lineshapes in titanium oxides: Influence of valence-band profile and of Coster-Kronig processes, Phys. Rev. B. 69 (2004) 155421. https://doi.org/10.1103/PhysRevB.69.155421.

[58]  R. Nyholm, N. Martensson, A. Lebugle, U. Axelsson, Auger and Coster-Kronig broadening effects in the 2p and 3p photoelectron spectra from the metals [22] Ti- [30] Zn, J. Phys. F: Met. Phys. 11 (1981) 1727–1733. https://doi.org/10.1088/0305-4608/11/8/025.

[59]  D. Jaeger, J. Patscheider, A complete and self-consistent evaluation of XPS spectra of TiN, Journal of Electron Spectroscopy and Related Phenomena. 185 (2012) 523–534. https://doi.org/10.1016/j.elspec.2012.10.011.

[60]  F.E. Oropeza, I.J. Villar-Garcia, R.G. Palgrave, D.J. Payne, A solution chemistry approach to epitaxial growth and stabilisation of $Bi_2Ti_2O_7$ films, J. Mater. Chem. A. 2 (2014) 18241–18245. https://doi.org/10.1039/C4TA04352A.

[61]  M. Demeter, M. Neumann, W. Reichelt, Mixed-valence vanadium oxides studied by XPS, Surface Science. 454–456 (2000) 41–44. https://doi.org/10.1016/S0039-6028(00)00111-4.

[62]  G. Silversmit, D. Depla, H. Poelman, G.B. Marin, R. De Gryse, Determination of the V2p XPS binding energies for different vanadium oxidation states (V5+ to V0+), Journal of Electron Spectroscopy and Related Phenomena. 135 (2004) 167–175. https://doi.org/10.1016/j.elspec.2004.03.004.

[63]  F. Ureña-Begara, A. Crunteanu, J.-P. Raskin, Raman and XPS characterization of vanadium oxide thin films with temperature, Applied Surface Science. 403 (2017) 717–727. https://doi.org/10.1016/j.apsusc.2017.01.160.

[64]  J. Mendialdua, R. Casanova, Y. Barbaux, XPS studies of V2O5, V6O13, VO2 and V2O3, Journal of Electron Spectroscopy and Related Phenomena. 71 (1995) 249–261. https://doi.org/10.1016/0368-2048(94)02291-7.

[65]  Shannon Radii, (n.d.). http://abulafia.mt.ic.ac.uk/shannon/ptable.php (accessed July 3, 2021).

[66]  M. Kowalik, R. Zalecki, A. Kołodziejczyk, Electronic States of Collosal Magnetoresistive Manganites $La_{0.67}Pb_{0.33}Mn_{1-x}Fe_xO_3$ from Photoemission Spectroscopy, Acta Phys. Pol. A. 117 (2010) 277–280. https://doi.org/10.12693/APhysPolA.117.277.

[67]  A.K. Shukla, P. Krüger, R.S. Dhaka, D.I. Sayago, K. Horn, S.R. Barman, Understanding the 2 p core-level spectra of manganese: Photoelectron spectroscopy experiments and Anderson impurity model calculations, Phys. Rev. B. 75 (2007) 235419. https://doi.org/10.1103/PhysRevB.75.235419.

[68]  P.S. Bagus, C.J. Nelin, C.R. Brundle, N. Lahiri, E.S. Ilton, K.M. Rosso, Analysis of the Fe 2p XPS for hematite $α Fe_2O_3$ : Consequences of covalent bonding and orbital splittings on multiplet splittings, J. Chem. Phys. 152 (2020) 014704. https://doi.org/10.1063/1.5135595.

[69]  E.G. Rini, M.K. Gupta, R. Mittal, A. Mekki, M.H. Al Saeed, S. Sen, Structural change from Pbnm to R3̄c phase with varying Fe/Mn content in (1-x) LaFeO3.xLaMnO3 solid solution leading to modifications in octahedral tilt and valence states, Journal of Alloys and Compounds. 883 (2021) 160761. https://doi.org/10.1016/j.jallcom.2021.160761.

[70]  M.C. Biesinger, B.P. Payne, A.P. Grosvenor, L.W.M. Lau, A.R. Gerson, R.St.C. Smart, Resolving surface chemical states in XPS analysis of first row transition metals, oxides and hydroxides: Cr, Mn, Fe, Co and Ni, Applied Surface Science. 257 (2011) 2717–2730. https://doi.org/10.1016/j.apsusc.2010.10.051.

[71]  Ch. Rayssi, S. El.Kossi, J. Dhahri, K. Khirouni, Frequency and temperature-dependence of dielectric permittivity and electric modulus studies of the solid solution $Ca_{0.85}Er_{0.1}Ti_{1-x}Co_{4x/3}O_3$ $(0 \leq x \leq 0.1)$, RSC Adv. 8 (2018) 17139–17150. https://doi.org/10.1039/C8RA00794B.

[72]  M. Abdullah Dar, K. Majid, K.M. Batoo, R.K. Kotnala, Dielectric and impedance study of polycrystalline Li0.35−0.5Cd0.3Ni Fe2.35−0.5O4 ferrites synthesized via a citrate-gel auto combustion method, Journal of Alloys and Compounds. 632 (2015) 307–320. https://doi.org/10.1016/j.jallcom.2015.01.190.

[73]  W. Cao, Calculation of various relaxation times and conductivity for a single dielectric relaxation process, Solid State Ionics. 42 (1990) 213–221. https://doi.org/10.1016/0167-2738(90)90010-O.

[74] D.C. Sinclair, A.R. West, Impedance and modulus spectroscopy of semiconducting BaTiO$_3$ showing positive temperature coefficient of resistance, Journal of Applied Physics. 66 (1989) 3850–3856. https://doi.org/10.1063/1.344049.

[75] R. Gerhardt, Impedance and dielectric spectroscopy revisited: Distinguishing localized relaxation from long-range conductivity, Journal of Physics and Chemistry of Solids. 55 (1994) 1491–1506. https://doi.org/10.1016/0022-3697(94)90575-4.

[76] B.K. Barick, K.K. Mishra, A.K. Arora, R.N.P. Choudhary, D.K. Pradhan, Impedance and Raman spectroscopic studies of (Na$_{0.5}$ Bi$_{0.5}$ )TiO$_3$, J. Phys. D: Appl. Phys. 44 (2011) 355402. https://doi.org/10.1088/0022-3727/44/35/355402.

[77] D.K. Pradhan, R.N.P. Choudhary, C. Rinaldi, R.S. Katiyar, Effect of Mn substitution on electrical and magnetic properties of Bi0.9La0.1FeO3, Journal of Applied Physics. 106 (2009) 024102. https://doi.org/10.1063/1.3158121.

[78] H.S. Mohanty, A. Kumar, B. Sahoo, P.K. Kurliya, D.K. Pradhan, Impedance spectroscopic study on microwave sintered (1 − x) Na0.5Bi0.5TiO3−x BaTiO3 ceramics, J Mater Sci: Mater Electron. 29 (2018) 6966–6977. https://doi.org/10.1007/s10854-018-8683-2.

[79] P.S. Anantha, K. Hariharan, ac Conductivity analysis and dielectric relaxation behaviour of NaNO3–Al2O3 composites, Materials Science and Engineering: B. 121 (2005) 12–19. https://doi.org/10.1016/j.mseb.2004.12.005.

[80] W.G. Wang, X.Y. Li, Impedance and dielectric relaxation spectroscopy studies on the calcium modified Na$_{0.5}$ Bi$_{0.44}$ Ca$_{0.06}$ TiO$_{2.97}$ ceramics, AIP Advances. 7 (2017) 125318. https://doi.org/10.1063/1.5012108.

[81] J.T.S. Irvine, D.C. Sinclair, A.R. West, Electroceramics: Characterization by Impedance Spectroscopy, Adv. Mater. 2 (1990) 132–138. https://doi.org/10.1002/adma.19900020304.

[82] R. Amin, K. Samantaray, E.G. Rini, I. Bhaumik, S. Sen, Grain and grain boundary contributions to AC conductivity in ferroelectric Ba0.75Pb0.25Ti1-Zr O3 ceramics, Ceramics International. 47 (2021) 13118–13128. https://doi.org/10.1016/j.ceramint.2021.01.176.

[83] X.X. Huang, X.G. Tang, X.M. Xiong, Y.P. Jiang, Q.X. Liu, T.F. Zhang, The dielectric anomaly and pyroelectric properties of sol–gel derived (Pb,Cd,La)TiO3 ceramics, J Mater Sci: Mater Electron. 26 (2015) 3174–3178. https://doi.org/10.1007/s10854-015-2814-9.

[84] C. Ang, Z. Yu, L.E. Cross, Oxygen-vacancy-related low-frequency dielectric relaxation and electrical conduction in B i : S r T i O 3, Phys. Rev. B. 62 (2000) 228–236. https://doi.org/10.1103/PhysRevB.62.228.

[85] W. Cao, W. Li, Y. Feng, T. Bai, Y. Qiao, Y. Hou, T. Zhang, Y. Yu, W. Fei, Defect dipole induced large recoverable strain and high energy-storage density in lead-free Na$_{0.5}$ Bi$_{0.5}$ TiO$_3$ -based systems, Appl. Phys. Lett. 108 (2016) 202902. https://doi.org/10.1063/1.4950974.

[86] D.S. Kim, B.C. Kim, S.H. Han, H.-W. Kang, J.S. Kim, C.I. Cheon, Direct and indirect measurements of the electro-caloric effect in (Bi,Na)TiO$_3$ -SrTiO$_3$ ceramics, Journal of Applied Physics. 126 (2019) 234101. https://doi.org/10.1063/1.5117773.

[87] B. Li, J.B. Wang, X.L. Zhong, F. Wang, Y.K. Zeng, Y.C. Zhou, The coexistence of the negative and positive electrocaloric effect in ferroelectric thin films for solid-state refrigeration, EPL. 102 (2013) 47004. https://doi.org/10.1209/0295-5075/102/47004.

[88] F. Zhuo, Q. Li, Q. Yan, Y. Zhang, H.-H. Wu, X. Xi, X. Chu, W. Cao, Temperature induced phase transformations and negative electrocaloric effect in (Pb,La)(Zr,Sn,Ti)O$_3$ antiferroelectric single crystal, Journal of Applied Physics. 122 (2017) 154101. https://doi.org/10.1063/1.4986849.

[89] Y. Zhang, W. Li, Z. Wang, Y. Qiao, H. Xia, R. Song, Y. Zhao, W. Fei, Perovskite Sr $_{1−x}$ (Na$_{0.5}$ Bi$_{0.5}$ )$_x$ Ti$_{0.99}$ Mn$_{0.01}$ O$_3$ Thin Films with Defect Dipoles for High Energy-Storage and Electrocaloric Performance, ACS Appl. Mater. Interfaces. 11 (2019) 37947–37954. https://doi.org/10.1021/acsami.9b14815.